\documentclass[a4paper,11pt]{article}
\usepackage{graphicx}
\usepackage{here}
\usepackage{amsmath,amssymb,amsthm}
\usepackage{mathrsfs}
\usepackage{ascmac}
\usepackage[sort, comma]{natbib}
\usepackage[subrefformat=parens]{subcaption}
\usepackage{url}

\newtheorem{theo}{Theorem}
\newtheorem{lem}{Lemma}
\newcommand{\ol}[1]{\overline{#1}}
\newcommand{\cl}[1]{\mathcal{#1}}

\title{{\Large A continuous space model of new economic geography with a quasi-linear log utility function}}
\author{Kensuke Ohtake\thanks{Center for General Education, Shinshu University, Matsumoto, Nagano 390-8621, Japan,
E-mail: k\_ohtake@shinshu-u.ac.jp}}
\date{August 28, 2023}

\begin{document}
\maketitle

\begin{abstract}
We consider the extension of a tractable NEG model with a quasi-linear log utility to continuous space, and investigate the behavior of its solution mathematically. The model is a system of nonlinear integral and differential equations describing the market equilibrium and the time evolution of the spatial distribution of population density. A unique global solution is constructed and a homogeneous stationary solution with evenly distributed population is shown to be unstable. Furthermore, it is shown numerically that the destabilized homogeneous stationary solution eventually forms spiky spatial distributions. The number of the spikes decreases as the preference for variety increases or the transport cost decreases.

\end{abstract} 

\noindent
{\bf Keywords:\hspace{3mm}}new economic geography~\textperiodcentered~continuous racetrack economy~\textperiodcentered~self-organization~\textperiodcentered~spatial patterns~\textperiodcentered~number of cities


\newpage

\section{Introduction}
The core-periphery model proposed by \citet{Krug91} is a fundamental model in new economic geography. Many mathematical studies have been conducted on the model and its extensions (Various theoretical studies based on the core-periphery model are detailed in \citet{FujiKrugVenab}.). However, studying the behavior of their solutions is still difficult due to the strong nonlinearity of those models especially in the case of more than three multiple regions. Meanwhile, some mathematical models that are more tractable to analytical study have been proposed (See \citet{OttaTabThi}, \citet{ForsOtta} for example. \citet{FujiThis}, \citet{SatoTabuYama} and \citet{ZenTaka} systematically introduce the mathematical models in the field including such tractable ones.). Among such tractable models, \citet{Pfl} and \citet{GasCasCorr} are particularly useful in this study. \citet{Pfl} has greatly simplified the core-periphery model by using a quasi-linear log utility function of consumers. \citet{GasCasCorr} have extended Pfl\"{u}ger's model to the discrete multi-region case. \citet{AkaTakaIke} also consider the same quasi-linear log utility model in a discrete space but with consumer's idiosyncratic taste for location in migration behavior.

In this paper, we discuss the extension of \citet{Pfl} and \citet{GasCasCorr} to continuous space. Firstly, we construct the model in continuous space. This could facilitate theoretical studies of spatial economy consisting of so many regions that they can be considered a continuum\footnote{\citet{FujiKrugVenab} have introduced the core-periphery model in a one-dimensional periodic continuous space. An early analytical treatment of the original core-periphery model in a continuous space has been given by \citet{TabaEshiSakaTaka}. \citet{TabaEshi23} develop an analytical method for the continuous space model.}. We propose the model on $n$-dimensional space, but the concrete behavior of the solutions is considered on one-dimensional continuous periodic space (continuous racetrack model). The continuous racetrack model would be convenient for investigating the self-organization of spatial structures under completely symmetric conditions where addressable regions are not predetermined. Secondly, we construct a unique global solution. Thirdly, we show that a homogeneous steady-state solution of the racetrack model is unstable, so the flat-earth distribution is not sustainable. Fourthly, in the racetrack model, we show numerically that the density of mobile workers almost completely agglomerate to a finite number of regions that should be called ``cities", and that the number of the cities thus formed increases with the transport cost increases or the preference for variety weakens. 

We discuss the emergence of agglomeration in the form of bifurcation from the homogeneous solution. As \citet[p.398]{Ioann2012} states, such view of ``symmetry breaking" has played an important role in understanding spatial economy. Early examples of such studies have been given by \citet{PapaSmi83}, \citet{Matsu95}, and \citet{Matsu96}. The method used in this paper, based on an orthogonal function expansion of a solution of a linearized model in a continuous space, has been introduced by \citet[Chapter 6]{FujiKrugVenab} into the racetrack model. \citet{ChiAsh08} have further developed the method not only for the racetrack model, but even for the model on the two-dimensional sphere.

The rest of the paper is organized as follows. Section 2 derives the model. Section 3 constructs the unique global solution. Section 4 investigates the stability of the homogeneous stationary solution. Section 5 shows numerical results on the large-time behavior of the solutions of the racetrack model. Section 6 concludes.

\section{The model}
In this section, we introduce the model which we handle in this paper. We first review the Dixit-Stiglitz framework (\citet{DS77}). Next, we apply it to construct a model on continuous space. 

\subsection{Dixit-Stiglitz framework}
As in \citet{Pfl}, we use use the quasi-linear utility function of consumers defined as
\begin{equation}\label{quasi-lin-U}
U = \mu\ln M + A,~~\mu\in[0,1).
\end{equation}
Here, $M$ and $A$ represent a composite index of the consumption of manufacturing goods, and the consumption of the agricultural good, respectively. The composite index $M$ is defined by
\begin{equation}\label{M}
M = \left[\int_0^n q(i)^{\frac{\sigma-1}{\sigma}}di\right]^{\frac{\sigma}{\sigma-1}}, ~~\sigma\in(1,\infty),
\end{equation}
where $q(i)$ represents the consumption of a variety $i\in[0,n]$ of manufactured goods. The parameter $\sigma>1$ represents the elasticity of substitution between any two varieties. Note that the closer the value of $\sigma$ is to $1$, the stronger the consumer's preference for variety becomes. Let the agricultural good be the num\'{e}raire, then the budget constraint of each consumer is
\begin{equation}\label{bud-const}
A+\int_0^n p(i)q(i)di = Y,
\end{equation}
where $p(i)$ and $Y$ represent the price of a variety $i$ and the consumer's income, respectively. By maximizing the utility function \eqref{quasi-lin-U} under \eqref{M} and \eqref{bud-const}, we see that 
\begin{equation}\label{dem-j}
q(j) = \mu\frac{p(j)^{-\sigma}}{G^{1-\sigma}},
\end{equation}
for each $j\in[0,n]$. Here, $G$ denotes price index for manufacturing goods and it satisfies
\begin{equation}\label{price-index}
G^{1-\sigma}=\int_0^n p(i)^{1-\sigma}di.
\end{equation}
By using \eqref{M}, \eqref{bud-const}, \eqref{dem-j} and \eqref{price-index}, we obtain 
\[
M=\frac{\mu}{G},~~A=Y-\mu.
\]
Here, we assume that $A>0$ for all consumers. A sufficient condition for this assumption is discussed later in Subsection \ref{posagr}. Finally, the indirect utility of the consumer is
\[
V = \mu(\ln \mu - 1) + Y -\mu \ln G.
\]

\subsection{Extension to continuous space}
Let us assume that the economic regions are continuously distributed on a measurable set $\cl M$ in $\mathbb{R}^n$, where $n\in \mathbb{N}$. In the following, the regions are indexed by continuous variables $x,y\in\cl{M}$. Let $\phi(x)$ be the population density of immobile workers at a region $x\in \cl{M}$, and $\lambda(x)$ be the population density of mobile workers at $x\in \cl{M}$. Here, the density functions $\phi(x)$ and $\lambda(x)$ are such that the integrals $\int_{\cl A}\phi(x)dx$ and $\int_{\cl A}\lambda(x)dx$ give the immobile and mobile population on any subset $\cl A\subset \cl M$, respectively. The total population of the immobile and mobile workers are given by $\Phi$ and $\Lambda$, respectively: 
\[
\int_{\cl{M}}\phi(x)dx=\Phi,~~\mbox{and}~~\int_{\cl{M}}\lambda(x)dx=\Lambda.
\]
All manufactured goods produced in the same region $x$ are assumed to have the same price $p(x)$. As in \citet{Pfl}, we assume the iceberg transport cost, that is, we have to ship $T(x, y)\geq 1$ units of manufacturing good from a region $x\in \cl{M}$ in order to deliver one unit of it to a region $y \in \cl{M}$. Then, the price at $y$ of manufactured goods produced at $x$, denoted by $p(x, y)$, is given by $p(x, y)=p(x) T(x, y)$. We assume that the function $T$ is continuous and bounded on $\cl{M}\times\cl{M}$ satisfying 
\begin{equation}\label{ulbT}
1=T_1 \leq T(x,y) \leq T_2
\end{equation}
for any $x,y\in \cl{M}$. Due to the transport cost, the price indices can take different values in different regions. By \eqref{price-index}, the price index at $x$ denoted by $G(x)$ satisfies 
\begin{equation}\label{pind}
G(x)^{1-\sigma}=\int_\cl{M} n(y) (p(y)T(x, y))^{1-\sigma}dy,
\end{equation}
where $n(x)$ denotes the density of the number of available varieties of manufactured goods at $x$. Here, the integral is over the manifold $\cl{M}$, and thus $dy$ is the volume element of the manifold. Note that each manufacturing firm at region $x$ has to ship $T(x,y)$ times amount of demand in each region $y$, then the output $q(x)$ of the firm at $x$ is calculated by using \eqref{dem-j} as
\begin{equation}\label{qr}
\begin{aligned}
q(x) &= \mu\int_\cl{M} \left[\frac{p(x, y)^{-\sigma}}{G(y)^{1-\sigma}}\times T(x, y)\times \left(\phi(y) + \lambda(y)\right)\right]dy\\
&=\mu p(x)^{-\sigma}\int_\cl{M} T(x, y)^{1-\sigma}\left(\phi(y)+\lambda(y)\right)G(y)^{\sigma-1}dy.
\end{aligned}
\end{equation}
Let us consider firms' pricing behavior. All manufacturing firms producing the quantity $q(x)$ at $x$ face the same costs that are expressed by
\begin{equation}\label{prtech}
c_f + c_m q(x)
\end{equation}
where $c_f$ and $c_m$ is the fixed and marginal costs, respectively. Then, the profit earned by each firm at $x$ is
\begin{equation}\label{profit}
\Pi(x) = p(x)q(x) -\left(c_f+c_m q(x)\right).
\end{equation}
The firms are assumed to maximize their profits. The first-order condition of optimality $\frac{d\Pi(x)}{dp(x)}=0$ yields
\begin{equation}\label{prcm}
p(x) = \frac{\sigma}{\sigma-1}c_m.
\end{equation}
Putting \eqref{prcm} into \eqref{profit}, we see that
\[
\Pi(x) = \frac{c_m q(x)}{\sigma-1}-c_f.
\]
Due to the zero profit condition and \eqref{prcm}, we see that
\begin{equation}\label{freeentry}
c_f = \frac{1}{\sigma}p(x) q(x).
\end{equation}

We follow the assumption in \citet{ForsOtta}. That is, firms need $F$ units of mobile workers as the fixed input and $(\sigma-1)/\sigma$ units of immobile workers as the marginal inputs, i.e., $c_{f}=Fw(x)$ and $c_{m}=(\sigma-1)/\sigma$ in \eqref{prtech}. Here, $w(x)$ denotes nominal wage for mobile workers at $x$. The nominal wage for immobile workers is assumed to be fixed to $1$. Then, by \eqref{prcm} and \eqref{freeentry}, we obtain 
\begin{eqnarray}
&~&p(x) = 1, \label{pr1}\\
&~&q(x) = \sigma F w(x). \label{qrsFwr}
\end{eqnarray}
\eqref{pind}, \eqref{qr}, \eqref{pr1} and \eqref{qrsFwr} lead to
\begin{align}
& G(x)^{1-\sigma} =\int_{\cl{M}} n(y) T(x, y)^{1-\sigma}dy, \label{gn}\\
&w(x)=\frac{\mu}{\sigma F}\int_{\cl{M}} T(x, y)^{1-\sigma}\left(\phi(y)+\lambda(y)\right)G(y)^{\sigma-1}dy.\label{wn}
\end{align}
By assuming that a single manufacturing firm is engaged only in the production of a single variety, we see that
\[
n(x) = \frac{\lambda(x)}{F}.
\]
The real wage of mobile workers at $x$ is defined by 
\begin{equation}\label{realwage}
\omega(x) = w(x) - \mu\ln G(x),
\end{equation}
and the average real wage is defined by $\frac{1}{\Lambda}\int_{\cl{M}} \omega(y) \lambda(y)dy$. 

The mobile population flows out of regions with lower-than-average real wages and into regions with higher-than-average real wages. The size of the outflow/inflow population is proportional to the difference between the real wage and the average real wage. Then, the temporal change in the mobile population density at each region is assumed to be governed by the differential equation in time
\begin{equation}\label{dtpop}
\frac{\partial}{\partial t}\lambda(x) = \gamma\left[\omega(x)-\frac{1}{\Lambda}\int_{\cl{M}} \omega(y) \lambda(y)dy\right]\lambda(x),
\end{equation}
where $\gamma>0$ represents an adjustment speed of agglomeration. Then, note that the conservation of population $\frac{d}{dt}\Lambda = 0$ holds. One might view this equation as representing the macroscopic dynamics of the population distribution, i.e., the average dynamics of migration behavior by microscopic individual players, including uncertainty and randomness.

The entire system of these equations \eqref{gn}-\eqref{dtpop}, with the explicit statement that the functions $G, w, \omega$, and $\lambda$ depend not only on the spatial variable $x\in\cl{M}$ but also on the time variable $t\in[0,\infty)$, can be shown as follows.
\begin{equation}\label{1}
\left\{
\begin{aligned}
&G(t, x) = \left[\frac{1}{F}\int_{\cl{M}} \lambda(t, y)T(x, y)^{1-\sigma}dy\right]^{\frac{1}{1-\sigma}},\\
&w(t, x)=\frac{\mu}{\sigma F}\int_{\cl{M}} T(x, y)^{1-\sigma}\left(\phi(y)+\lambda(t, y)\right)G(t, y)^{\sigma-1}dy,\\ 
&\omega(t, x) = w(t, x)-\mu\ln G(t, x),\\
&\frac{\partial \lambda(t,x)}{\partial t} = \gamma\left[\omega(t,x) - \frac{1}{\Lambda}\int_{\cl{M}} \omega(t,y)\lambda(t,y)dy\right]\lambda(t,x)
\end{aligned}\right.
\end{equation}
with an initial condition $\lambda(0, x)=\lambda_0(x)$. 

In the following, we investigate the properties of the model \eqref{1} in detail, on a  one-dimensional circumference with radius $r>0$ denoted by $S$. In addition, the transportation cost function $T(x, y)$ is assumed to be 
\begin{equation}\label{rtT}
T(x, y) = e^{\tau|x-y|},~\tau>0
\end{equation}
where $|x-y|$ means the shorter distance between $x$ and $y$ on $S$.  Here, $|x-y|=|y-x|$ holds for any $x,y\in S$, so the function $T(x,y)$ is symmetric with respect to the variables on $S$; $T(x,y)=T(y,x)$. Moreover, the immobile population density is assumed to be homogeneous;
\begin{equation}\label{phih}
\phi(x)\equiv \ol{\phi}=\Phi/(2\pi r)~\mbox{on}~S.
\end{equation}
In the following of the paper, we simply refer to \eqref{1} on $S$ with the conditions \eqref{rtT} and \eqref{phih} as {\it racetrack model}. In the analysis of the racetrack model, the crucial parameters $\sigma$ and $\tau$ frequently appear in the form of multiplication by $(\sigma-1)\tau$. Therefore, it is useful to introduce the variable
\[
\alpha = (\sigma-1)\tau
\]
to improve the following discussion's perspective. By definition, $\alpha \geq 0$.

\subsection{Notes on specific calculation on $S$}\label{subsec:notesoncomputation}

\subsubsection{Identification of a function on $S$ with that on $[-\pi,\pi]$}
A point $x\in S$ corresponds one-to-one to a point $\theta\in[-\pi,\pi]$, so
\[
x = x(\theta).
\]
Therefore, the function $f$ on $S$ can be written as 
\[
f(x)=f\left(x(\theta)\right)=:\tilde{f}(\theta),
\]
by a periodic function $\tilde{f}$ on $[-\pi, \pi]$.  Note that the line element $dx$ of $S$ is expressed by $d\theta$ as $dx = rd\theta$. Then, the integral over $S$ of the function $f$ is calculated by 
\[
\begin{aligned}
\int_S f(x)dx &= \int_{-\pi}^\pi f\left(x(\theta)\right)rd\theta \\
&= \int_{-\pi}^\pi \tilde{f}(\theta)rd\theta.
\end{aligned}
\]
In the following, we just identify the function $\tilde{f}$ with the corresponding function $f$ and denote it as $f$ simply, when no confusion can arise.

\subsubsection{Distance function}
The distance $|x-y|$ between $x(\theta^\prime),~y(\theta)\in S$ is computed as 
\[
|x-y| = r\times \min\left\{ |\theta^\prime-\theta|_{{\rm abs}}, 2\pi-|\theta^\prime-\theta|_{{\rm abs}} \right\},
\]
where $|\cdot|_{\rm abs}$ is a function that returns the absolute value of its argument.

\subsection{Sufficient condition for positive agricultural demand}\label{posagr}
We assume that $A=Y-\mu>0$. This assumption always holds as for the immobile workers because their nominal wages are $1>\mu$. On the other hand, the following theorem claims that a large enough immobile population guarantees that the mobile workers' nominal wage to be greater than $\mu$.
\begin{theo}
In the racetrack model, if 
\begin{equation}\label{positive_agr}
\Phi > \frac{\sigma\Lambda\alpha\pi r}{1-e^{-\alpha r \pi}}
\end{equation}
holds, then $w(x)>\mu$ for any $x\in S$. 
\end{theo}
\begin{proof}
In the racetrack model, the nominal wage of the mobile workers at $x$ is given by
\begin{equation}\label{nom_racetrack}
w(x) = \frac{\mu}{\sigma F}\int_S[\ol{\phi}+\lambda(y)]G(y)^{\sigma-1}e^{-\alpha|x-y|}dy.
\end{equation}
An inequality
\[
G(x)^{\sigma-1} > \frac{F}{\Lambda}.
\]
follows from
\[
G(x)^{1-\sigma} = \frac{1}{F}\int_S\lambda(y)e^{-\alpha|x-y|}dy<\frac{\Lambda}{F}.
\]
Then, we obtain
\begin{eqnarray}
w(x) &=& \frac{\mu}{\sigma F}\int_S[\ol{\phi}+\lambda(y)]G(y)^{\sigma-1}e^{-\alpha|x-y|}dy \nonumber \\ 
&>& \frac{\mu}{\sigma\Lambda}\int_S[\ol{\phi}+\lambda(y)]e^{-\alpha|x-y|}dy \nonumber \\
&>& \frac{\mu\Phi}{\sigma\Lambda2\pi r} \int_S e^{-\alpha|x-y|}dy = \frac{\mu\Phi(1-e^{-\alpha r\pi})}{\sigma\Lambda\alpha\pi r}.\label{wlowestm}
\end{eqnarray}
Here, we use
\begin{equation}\label{intoftenused}
\int_S e^{-\alpha|x-y|}dy = \frac{2(1-e^{-\alpha\pi r})}{\alpha},
\end{equation}
which follows from the techniques for calculation in Subsection \ref{subsec:notesoncomputation}. Applying \eqref{wlowestm} to \eqref{nom_racetrack}, we see that $w(x)>\mu$ follows from \eqref{positive_agr}.

\end{proof}

\section{Existence results}
In this section, the model \eqref{1} is formulated as an ordinary differential equation in a Banach space. Then, we construct a solution of it applying Picard–Lindel\"{o}f theorem. We also see that the solution depends continuously on the initial value.

\subsection{Mathematical formulation}
We formulate \eqref{1} as an initial value problem of an ordinary differential equation in a Banach space.
\subsubsection{Basic function spaces}
Let $L^1(\cl M)$ be the Banach space of all measurable functions $f$ on $\cl M$ which satisfy $\left\|f\right\|_{L^1}:=\int_{\cl M}|f(x)|dx<\infty$. We introduce a closed subset of $L^1(\cl M)$ given by
\[
L^1_{\Lambda}(\cl M):=\left\{f\in L^1(\cl M)\middle| f\geq 0~{\rm a.e.}~x\in \cl{M}, \int_{\cl M}f(x)dx=\Lambda\right\}.
\]
Let $C_b(\cl M)$ be the Banach space of all bounded continuous functions $f$ on $\cl M$ equipped with the norm $\left\|f\right\|_\infty:=\sup_{x\in \cl M}\left|f(x)\right|$. We intend to construct a solution $\lambda$, $G$, $w$, and $\omega$ of \eqref{1} such that $\lambda(t, \cdot)\in L^1_{\Lambda}(\cl{M})$, $G(t,\cdot)\in C_b(\cl M)$, $w(t,\cdot)\in C_b(\cl M)$, and $\omega(t,\cdot)\in C_b(\cl M)$ in each time $t\geq 0$.

\subsubsection{Operators on Banach spaces}
We define operators $G$, $w$, and $\omega$ that map $L^1(\cl{M})$ to $C_b(\cl{M})$ by
\begin{align}
&G(\lambda)(x) = \left[\frac{1}{F}\int_{\cl{M}} \left|\lambda(y)\right|T(x, y)^{1-\sigma}dy\right]^{\frac{1}{1-\sigma}}, \hspace{27.5mm}x\in \cl{M},\label{opG}\\
&w(\lambda)(x) = \frac{\mu}{\sigma F}\int_{\cl{M}} T(x, y)^{1-\sigma}\left(\phi(y)+\left|\lambda(y)\right|\right)G(\lambda)(y)^{\sigma-1}dy,~x\in \cl{M},\label{opw}\\
&\omega(\lambda)(x) = w(\lambda)(x)-\mu\ln G(\lambda)(x),\hspace{43.3mm}x\in \cl{M},\label{opom}
\end{align}
which correspond to the price index, nominal wage, and real wage in \eqref{1}, respectively. Note that in \eqref{opG} and \eqref{opw}, it is $\left|\lambda(y)\right|$ , not $\lambda(y)$. By using these operators, we define an operator $\Psi:L^1(\cl M)\to L^1(\cl{M})$ by
\begin{equation}\label{opPsi}
\Psi(\lambda)(x) = \gamma\left[\omega(\lambda)(x)-\frac{1}{\Lambda}\int_\cl{M}\omega(\lambda)(y)\lambda(y)dy\right]\lambda(x),\hspace{5mm} x\in \cl{M}
\end{equation}
which corresponds to the right-hand side of the differential equation of \eqref{1}.

\subsubsection{ODE in Banach space}
Then, using the operator \eqref{opPsi}, we consider an initial value problem for an ordinary equation 
\begin{equation}\label{F1}
\left\{
\begin{aligned}
&\frac{d\lambda}{dt}(t) = \Psi(\lambda(t)),\\
&\lambda(0) = \lambda_0.
\end{aligned}\right.
\end{equation}
in $L^1(\cl{M})$. The initial function $\lambda_0$ is taken from $L^1_{\Lambda}(\cl{M})$. Let us define a solution $\lambda$ of \eqref{1} by a solution $\lambda$ of \eqref{F1} satisfying $\lambda(t)\in L^1_\Lambda$ for  $t\geq 0$ if it exists.

Let $X=C\left(I; B \right)$ be the space of all continuous functions 
\[
f: I \to B,
\]
where $I\subset\mathbb{R}$ and $B$ are an interval and a Banach space, respectively. If $I$ is a closed bounded interval, then $X$ is a Banach space equipped with the norm $\left\|\cdot\right\|_X$ defined by
\begin{equation}\label{Xnorm}
\left\|f\right\|_X := \max_{t\in I}\left\|f(t)\right\|_B,
\end{equation}
where $\left\|\cdot\right\|_B$ is the norm of $B$. Similarly, let $C^1\left(I; B \right)\subset X$ be the space of all continuously differentiable $B$-valued functions on $I$.

\subsection{Local solution}
The global existence and uniqueness of the solution of \eqref{F1} is attributed to the Lipschitz continuity and boundedness of $\Psi(\lambda)$ that are shown in the following theorems. Their proofs are given in Appendix.

Let $X=C\left([0,c]; L^1(\cl M)\right)$ under a certain value of $c>0$. We introduce the closed subset in $X$ by
\[
Q_{\lambda_0}:=\left\{\lambda\in X\middle|\left\|\lambda-\lambda_0\right\|_X\leq b\right\}
\]
for $\lambda_0\in L^1_\Lambda$. Here, $b\geq 0$ is assumed to satisfy $\Lambda - b > 0$.

To construct a local solution of \eqref{F1}, the following two theorems are essential. See Appendix for the proofs.
\begin{theo}\label{th:locapr}
There exists a constant $k>0$ such that
\[
\left\|\Psi(\lambda(t))\right\|_{L^1} \leq k
\]
for all $t\in [0, c]$, holds for any $\lambda\in Q_{\lambda_0}$.
\end{theo}

\begin{theo}\label{th:lip}
There eixts a constant $\cl{L}>0$ such that
\begin{equation}\label{lipPsi}
\begin{aligned}
&\|\Psi(\lambda_1(t)) - \Psi(\lambda_2(t))\|_{L^1}
\leq \cl{L}\|\lambda_1(t)-\lambda_2(t)\|_{L^1}
\end{aligned}
\end{equation}
for all $t\in [0, c]$, holds for any $\lambda_1,\lambda_2\in Q_{\lambda_0}$.
\end{theo}

In the same manner of the proof of the generalized Picard–Lindel\"{o}f theorem (See \citet[p.78]{ZeidlerFixedPoint}), Theorem \ref{th:locapr} and \ref{th:lip} enable us to construct a local solution as in the following theorem.
\begin{theo}\label{th:locsol}
For any $\lambda_0\in L^1_\Lambda(\cl M)$, there exists $c>0$ such that \eqref{F1} has a unique local solution $\lambda\in C^1\left([0,c];L^1_\Lambda(\cl M)\right)$.
\end{theo}

\noindent
{\bf Note}: That $\lambda$ is not only in $C$ but also in $C^1$ immediately follows from the fact that $\Psi(\lambda(t))$ is continuous with respect to $t$.

\noindent
{\bf Note}: It is noteworthy that the existence of the local solutions does not depend on the parameters $\sigma>1$, $\mu\in[0,1)$, $F>0$, $\Phi>0$, $\Lambda>0$, $T_1>0$, and $T_2>0$. Moreover, no additional constraints need to be placed on these parameters when constructing global solutions.

\vspace{5mm}
The next theorem states that the local solutions continuously depend on their initial values. It is obtained by the Lipschitz continuity of $\Psi$ and Gronwall's lemma\footnote{See \citet[pp.82-83]{ZeidlerFixedPoint}}.
\begin{theo}
\begin{equation*}
\left\|\lambda_1(t)-\lambda_2(t)\right\|_{L^1} \leq \left\|\lambda_1(0)-\lambda_2(0)\right\|_{L^1} e^{\cl{L}t}
\end{equation*}
for all $t\in[0,c]$, holds for any local solutions $\lambda_1,\lambda_2$. Here, $\cl{L}>0$ is the Lipschitz constant in Theorem \ref{th:lip}. 
\end{theo}

\subsection{Global solution}
We have  the following a priori estimate. See Appendix for the proof.
\begin{theo}\label{th:apr}
For any $\lambda\in C\left([0,\infty); L^1_\Lambda\right)$, there exists a constant $K>0$ such that
\[
\left\|\Psi\left(\lambda(t)\right)\right\|_{L^1} \leq K
\]
for all $t\in[0,\infty)$ holds.
\end{theo}

Hence, the local solution geven in Theorem \ref{th:locsol} is extended to a global solution by the a priori estimate\footnote{See \citet[pp.80-81]{ZeidlerFixedPoint}.} given in Theorem \ref{th:apr}. 
\begin{theo}
For any $\lambda_0\in L^1_\Lambda(\cl M)$, there exists a unique global solution $\lambda\in C^1\left([0,\infty);L^1_\Lambda(\cl M)\right)$ of \eqref{F1}.
\end{theo}

\section{Stationary solution}
In the section, we consider a homogeneous stationary solution of the racetrack model and investigate its stability. 

\subsection{Homogeneous stationary solution}
The racetrack model has a homogeneous stationary solution $\ol{\lambda}\equiv \Lambda/(2\pi r)$. In this case, the nominal wage $\ol{w}$ and price index $\ol{G}$ are also spatially homogeneous and satisfy
\[
\begin{aligned}
&\ol{w}=\frac{\mu(\ol{\phi}+\ol{\lambda})}{\sigma\ol{\lambda}},\\
&\ol{G}^{1-\sigma}=\frac{\ol{\lambda}}{F}\int_S e^{-\alpha|x-y|}dy=\frac{2\ol{\lambda}(1-e^{-\alpha r\pi})}{F\alpha},
\end{aligned}
\]
and then, the real wage $\ol{\omega}$ is also homogeneous. Here, note that $\ol{G}$ does not depend on $x\in S$ by \eqref{intoftenused}.

\subsection{Stability of the stationary solution}
\subsubsection{Linearized equations}
Let small perturbations added to the population density of mobile workers, nominal wage, price index, and real wage be $\Delta \lambda(x),\Delta w(x), \Delta G(x), \Delta \omega(x)$, respectively. Note that
\[
\int_S \Delta \lambda(x)dx=0.
\]
Substituting $\lambda(t,x)=\ol{\lambda}+\Delta \lambda(t,x),w(t,x)=\ol{w}+\Delta w(t,x),G(t,x)=\ol{G}+\Delta G(t,x)$ and $\omega(t,x)=\ol{\omega}+\Delta\omega(t,x)$ into \eqref{1}, and neglecting second and higher order terms such as $\Delta \lambda\Delta G$, we obtain the linearized system as the following: 
\begin{equation}\label{linsys}
\left\{
\begin{aligned}
&\Delta w(t,x) = \frac{\mu(\sigma -1)(\ol{\phi}+\ol{\lambda})\ol{G}^{\sigma-2}}{F\sigma}\int_S\Delta G(t,y)e^{-\alpha|x-y|}dy\\
&\hspace{50mm}+ \frac{\mu\ol{G}^{\sigma-1}}{F\sigma}\int_S\Delta \lambda(t,y)e^{-\alpha|x-y|}dy,\\
&\Delta G(t,x) = \frac{\ol{G}^\sigma}{(1-\sigma)F}\int_S\Delta \lambda(t,y)e^{-\alpha|x-y|}dy,\\
&\Delta\omega(t,x) = \Delta w(t,x) -\frac{\mu}{\ol{G}}\Delta G(t,x),\\
&\frac{\partial \Delta \lambda}{\partial t}(t,x) = \gamma\ol{\lambda}\Delta \omega(t,x).
\end{aligned}
\right.
\end{equation}
Note that $\int_S\Delta\omega(t,x)dx=0$ when linearizing the differential equation.  Let us define the Fourier coefficients $\hat{f}_n$ of a function $\tilde{f}(\theta), ~\theta\in[-\pi, \pi]$ identified with $f(x),~x\in S$.
\[
\hat{f}_n = \int_S \tilde{f}(\theta) e^{-in\theta} d\theta.
\]
Here, $i^2=-1$ and $n=0,\pm 1, \pm 2, \cdots$. Then, for $\theta^\prime\in[-\pi,\pi]$, the Fourier expansions of the small perturbations are\footnote{Here, the functions $\Delta \lambda$, $\Delta w$, $\Delta G$, and $\Delta \omega$ on $S$ are identified with the corresponding periodic functions $\Delta \tilde{\lambda}$, $\Delta \tilde{w}$, $\Delta \tilde{G}$, and $\Delta \tilde{\omega}$ on $[-\pi, \pi]$, respectively (See Subsection \ref{subsec:notesoncomputation}).}
\begin{eqnarray}
&~&\Delta \lambda(t,\theta^\prime) = \frac{1}{2\pi}\sum_n \hat{\lambda}_n(t) e^{in\theta^\prime},\label{dellamexp}\\
&~&\Delta w(t,\theta^\prime) = \frac{1}{2\pi}\sum_n \hat{w}_n(t) e^{in\theta^\prime},\\
&~&\Delta G(t,\theta^\prime) = \frac{1}{2\pi}\sum_n \hat{G}_n(t) e^{in\theta^\prime},\\
&~&\Delta\omega(t,\theta^\prime) = \frac{1}{2\pi}\sum_n \hat{\omega}_n(t) e^{in\theta^\prime},\label{delomexp}
\end{eqnarray}
where, $e^{in\theta^\prime}$ is the $n$-th Fourier mode.  Substituting \eqref{dellamexp}-\eqref{delomexp} into the linearized system \eqref{linsys} yields
\begin{align}\label{wf}
&\sum_n \hat{w}_n e^{in\theta^\prime} 
= \frac{\mu(\sigma -1)(\ol{\phi}+\ol{\lambda})\ol{G}^{\sigma-2}}{F\sigma}\int_{-\pi}^\pi \sum_n \hat{G}_n e^{in\theta} e^{-\alpha r|\theta^\prime-\theta|}rd\theta\nonumber \\ 
&\hspace{30mm} + \frac{\mu\ol{G}^{\sigma-1}}{F\sigma}\int_{-\pi}^\pi \sum_n \hat{\lambda}_n e^{in\theta} e^{-\alpha r|\theta^\prime-\theta|}rd\theta \nonumber \\ 
&\hspace{18mm} = \frac{\mu(\sigma -1)(\ol{\phi}+\ol{\lambda})\ol{G}^{\sigma-2}}{F\sigma}\sum_n \hat{G}_n E_n e^{in\theta^\prime} \nonumber \\
&\hspace{60mm} + \frac{\mu\ol{G}^{\sigma-1}}{F\sigma}\sum_n \hat{\lambda}_n E_n e^{in\theta^\prime},\\
& \sum_n \hat{G}_n e^{in\theta^\prime}
= \frac{\ol{G}^\sigma}{(1-\sigma)F}\sum_n \hat{\lambda}_n E_n e^{in\theta^\prime}, \label{gf} \\ 
& \sum_n \hat{\omega}_n e^{in\theta^\prime} 
= \sum_n \hat{w}_ne^{in\theta^\prime} - \frac{\mu}{\ol{G}} \sum_n \hat{G}_n e^{in\theta^\prime},\label{omf}
\end{align}
and 
\begin{equation}\label{dlamdtf}
\sum_n \frac{d}{dt}\left(\hat{\lambda}_n\right) e^{in\theta^\prime} = \gamma\ol{\lambda}\sum_n \hat{\omega}_n e^{in\theta^\prime}.
\end{equation}
In \eqref{wf} and \eqref{gf}, the coefficient $E_n$ is given as
\[
E_n  = \frac{2\alpha r^2\left(1-(-1)^{|n|}e^{-\alpha r \pi}\right)}{n^2 + \alpha^2r^2},
\]
and it satisfies $\int_{-\pi}^\pi e^{in\theta}e^{-\alpha r|\theta^\prime-\theta|}rd\theta=E_n e^{in\theta^\prime}$.

From \eqref{wf}-\eqref{dlamdtf} the following equations  
\begin{equation}\label{hatsys}
\begin{aligned}
&\hat{w}_n = \frac{\mu(\sigma -1)(\ol{\phi}+\ol{\lambda})\ol{G}^{\sigma-2}}{F\sigma}E_n\hat{G}_n + \frac{\mu\ol{G}^{\sigma-1}}{F\sigma}E_n\hat{\lambda}_n,\\
&\hat{G}_n = \frac{\ol{G}^\sigma}{(1-\sigma)F} E_n \hat{\lambda}_n,\\
&\hat{\omega}_n = \hat{w}_n  - \frac{\mu}{\ol{G}}\hat{G}_n,\\
&\frac{d}{dt}\hat{\lambda}_n = \gamma\ol{\lambda}\hat{\omega}_n
\end{aligned}
\end{equation}
for each $n$ are obtained. Solving the first three equation of \eqref{hatsys} gives 
\[
\hat{\omega}_n = 
\frac{\mu\ol{G}^{\sigma-1}E_n}{F}\left\{\frac{2\sigma-1}{\sigma(\sigma-1)}-\frac{(\ol{\phi}+\ol{\lambda})}{\sigma F}\ol{G}^{\sigma-1}E_n\right\}\hat{\lambda}_n
\]
for each $n$. Then, the fourth equation of \eqref{hatsys} becomes
\[
\begin{aligned}
\frac{d}{dt}\hat{\lambda}_n = \gamma\ol{\lambda}\cdot\frac{\mu\ol{G}^{\sigma-1}E_n}{F}\left\{\frac{2\sigma-1}{\sigma(\sigma-1)}-\frac{(\ol{\phi}+\ol{\lambda})}{\sigma F}\ol{G}^{\sigma-1}E_n\right\}\hat{\lambda}_n
\end{aligned}
\]
for each $n$. This means that if
\begin{equation}\label{eig}
J_n := \frac{\mu\ol{G}^{\sigma-1}E_n}{F}\left\{\frac{2\sigma-1}{\sigma(\sigma-1)}-\frac{(\ol{\phi}+\ol{\lambda})}{\sigma F}\ol{G}^{\sigma-1}E_n\right\}
\end{equation}
is positive (negative), then the $n$-th mode is unstable (stable). Obviously, since $\frac{\mu\ol{G}^{\sigma-1}E_n}{F}>0$, the sign of $J_n$ is determined by the value in the braces of \eqref{eig}.

\subsubsection{No black hole condition}
Let us consider a closed economy with infinitely weak preference for variety or infinitely high transportation cost. We can prove that when $\sigma\to\infty$ or $\tau\to\infty$, i.e., $\alpha\to\infty$,
\[
\ol{G}^{\sigma-1}E_n \to \frac{F}{\ol{\lambda}}=\frac{2\pi Fr}{\Lambda}.
\]
for any number $n$. Therefore, the coefficient \eqref{eig} satisfies 
\[
J_n \to \frac{\mu2\pi r}{\Lambda}\left\{\frac{2\sigma-1}{\sigma(\sigma-1)}-\frac{\Phi+\Lambda}{\sigma\Lambda}\right\}
\]
when $\alpha\to\infty$ for any number $n$. Here, we use $\ol{\lambda}=\frac{\Lambda}{2\pi r}$ and $\ol{\phi}=\frac{\Phi}{2\pi r}$. Therefore, if $\frac{\sigma}{\sigma-1}>\Phi/\Lambda$ holds, then $J_n>0$, so the agglomeration occurs in spite of closed economy. Then, to eliminate such a situation where the centripetal forces are too strong, so-called the no black hole condition given by
\begin{equation}\label{nbh}
\frac{\sigma}{\sigma-1}<\frac{\Phi}{\Lambda}
\end{equation}
must be satisfied.

\subsubsection{Infinite number of unstable modes}
As discussed above, for any number $n$, if the preference for variety is sufficiently weak or the transport cost is sufficiently high, the $n$-th Fourier mode can be stabilized as long as \eqref{nbh} holds. Meanwhile, no matter how weak the preference for variety is (or how high the transport cost is), if the wave number $n$ is sufficiently large, the Fourier modes with wave numbers above $n$ are unstable. This fact can be written as the following theorem.
\begin{theo}\label{instabilitytheorem}
For any $\sigma>1$ and any $\tau>0$, there is an integer $\tilde{n}$ such that $J_n>0$ for any $|n|>|\tilde{n}|$. 
\end{theo}
The proof is straightforward by using \eqref{eig}. In fact, 
\[
\begin{aligned}
J_n &=\frac{\mu\ol{G}^{\sigma-1}E_n}{F}
\left\{
\frac{2\sigma-1}{\sigma(\sigma-1)}
- \frac{\alpha^2 r^2(\ol{\phi}+\ol{\lambda})\left(1-(-1)^{|n|}e^{-\alpha r\pi})\right)}{\sigma\ol{\lambda}\left(n^2+\alpha^2 r^2\right)\left(1-e^{-\alpha r\pi}\right)}
\right\}.
\end{aligned}
\]
In the braces of the above, the second term converges to zero as $|n|\to \infty$, so one can see that $J_n >0$ for sufficiently large $|n|$ because the first term $(2\sigma-1)/(\sigma(\sigma-1))$ is positive. Obviously,  this proof makes use of the mathematical properties of the continuous space model. That is to say, in the continuous space model, there are an infinite number of the Fourier modes, so the wave number $n$ can be as large as desired.

Theorem \ref{instabilitytheorem} simply means that the homogeneous stationary solution $\ol{\lambda}=\Lambda/(2\pi r)$ is always unstable in the racetrack model.

\section{Numerical simulations}
In this section, we perform numerical computations of the racetrack model. The integral operator is discretized by the trapezoidal rule\footnote{Under the periodic boundary condition, applying the trapezoidal rule is equivalent to approximating the integral by a simple Riemann sum as described in \eqref{approxintegral} below.}. The differential equation is then discretized using the $4$-th order explicit Runge-Kutta method modified to satisfy the conservation of population. 

\subsection{Settings and overview}
Let us describe basic settings for the simulation. We identify $S$ with the interval $[-\pi,\pi]$ with mod $2\pi$. Then, the variable $x$ in the interval is discretized into $I$ nodal points $x_i=-\pi+(i-1)\Delta x$, $i=1,2,\cdots,I$, where $\Delta x = 2\pi/I$. Temporal variable $t\in[0,\infty)$ is discretized by $t_j=(j-1)\Delta t$, $j=1,2,\cdots$, with $\Delta t >0$. We set $I=256$ and $\Delta t = 0.01$ in our simulation\footnote{Since both time and space variables are real numbers in the model, they must be discretized by a sufficiently large number of nodes in the numerical computations. We think that the size between nodes should be on the order of $10^{-2}$ at most in both time and space, so we set $I=256$ (then $dx\fallingdotseq 0.025$) and $dt=0.01$.}. Let $f$ be a function on $[0,\infty)\times S$. Then, we denote $f(t_j,x_i)=f^j_i$ and $f^j = [f^j_1, f^j_2, \cdots, f^j_I]$. Note that approximating an integral $\int_{-\pi}^\pi f(t,x)dx$ by the trapezoidal rule, we obtain 
\begin{equation}\label{approxintegral}
\int_{-\pi}^\pi f(t,x)dx \approx \sum_{i=1}^I f(t, x_i)\Delta x.
\end{equation}

For an approximated solution $\lambda^j=[\lambda_1^j,\lambda_2^j,\cdots,\lambda_I^j]$, the explicit Runge-Kutta scheme that discretizes the differential equation gives a {\it candidate} for $\lambda^{j+1}$ denoted by $\nu^{j+1}$. Let us write this relationship as $\nu^{j+1}=f(\lambda^{j})$. Note that we do not adopt this value $\nu^{j+1}$ as $\lambda^{j+1}$, but instead, to satisfy the conservation law, we perform the normalization: $\lambda^{j+1}=\nu^{j+1}/\|\nu^{j+1}\|_1$, where $\|\nu^{j+1}\|_1$ denotes $\sum_{i=1}^I |\nu^{j+1}_i| \Delta x$. The normalized value $\lambda^{j+1}$ is then adopted as an approximation of the solution at the next point in time. In short, one step in the explicit numerical scheme is given by
\[
\begin{aligned}
&\nu^{j+1}=f(\lambda^{j}),\\
&\lambda^{j+1}=\nu^{j+1}/\|\nu^{j+1}\|_1
\end{aligned}
\]
for $j=0,1,2,\cdots$.

This explicit method gives a sequence of $I$-dimensional vectors $\lambda^0, \lambda^1, \lambda^2, \cdots$. The calculation is performed until the maximum norm $\|\lambda^{j+1}-\lambda^{j}\|_\infty$ becomes smaller than $\epsilon=10^{-10}$, and the numerical solution thus obtained is considered to be an approximation of a stationary solution $\lambda^*(x)$. Initial values $\lambda_0(x)$ are set by adding small perturbations to the homogeneous state $\ol{\lambda}\equiv 1/(2\pi)$. These small perturbations are randomly generated in each simulation. In conjunction with $\lambda^*(x)$, we also obtain an approximation of stationary realwage $\omega^*(x)$.

In the simulation, the elasticity of substitution $\sigma >1$ and the transportation cost coefficient $\tau>0$ are considered control parameters. Other parameters $\Lambda$, $\Phi$, $F$, $\mu$ and $\gamma$ are fixed at $\Lambda=1.0$, $\Phi=10.0$, $F=1.0$, $\mu=0.6$ and $\gamma=1.0$, respectively\footnote{For simplicity, we set $F=1.0$ and $\gamma=1.0$. The reason for setting $\Lambda=1.0$ and $\Phi=10.0$ is to satisfy condition \eqref{positive_agr} and \eqref{nbh}. In addition, $\mu=0.6$ is a value used in \citet[p.93]{FujiKrugVenab}.}. To guarantee the positive agricultural demand, we do not rely on the condition \eqref{positive_agr} but instead, we check that $w(x) > \mu~\forall x$ is satisfied at each time step.

The following figures \ref{fig:tau0.05}-\ref{fig:sig9.2} show the approximated stationary mobile population density $\lambda^*(x)$ and approximated stationary realwage $\omega^*(x)$ obtained in the above way\footnote{In the figures, the actual computed values are indicated by the blue dots. The dashed lines are just the interpolation for the plot.}. In any case, the stationary solutions are extremely non-uniform in space, having several spikes\footnote{The observation that the non-uniform (at least stable) stationary solutions are limited to spiky ones would be a robust property that does not depend on any particular parameter. In fact, in addition to the results presented below, numerical simulations have also been performed for $\mu=0.2$ and $\mu=0.4$, but none of them, including those presented here, led to non-spiky steady-state solutions.}. Moreover, we also see that the spiked regions enjoy the highest real wages compared to other regions. For each pair of control parameters $(\sigma, \tau)$, we perform several computations. Then, the number of spikes varied slightly depending on the initial distributions, so the largest number of the spikes in each simulation is referred to as the {\it maximum number} in the following. Additionally, it is noteworthy that the location of the spikes depends on the initial distributions much more sensitively than the number of the spikes.

\subsection{Change in transportation cost}
First, we focus on the effects of the transportation cost on the behavior of the solution of the racetrack model, so we only change $\tau$ here\footnote{We adopt $\sigma=5.0$ because this is the value used in \citet[p.93]{FujiKrugVenab}.}.

\begin{figure}[H]
 \begin{subfigure}{0.5\columnwidth}
  \centering
  \includegraphics[width=\columnwidth]{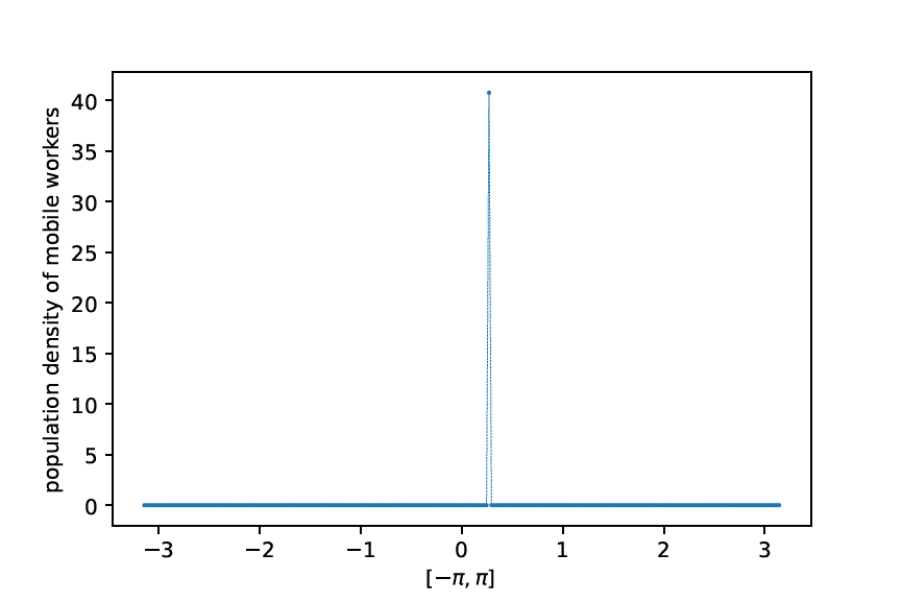}
  \caption{Mobile population density $\lambda^*$}
 \end{subfigure}
 \begin{subfigure}{0.5\columnwidth}
  \centering
  \includegraphics[width=\columnwidth]{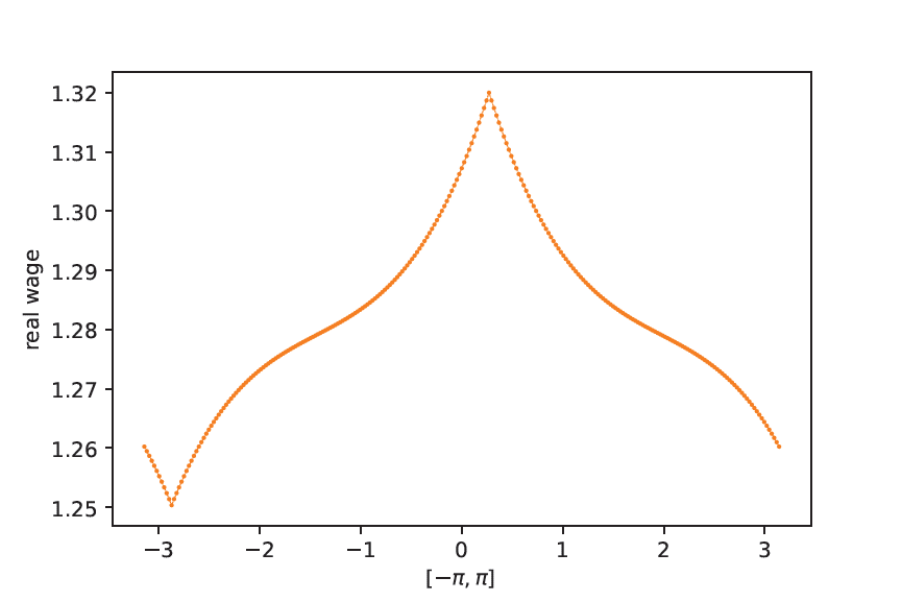}
  \caption{Real wage $\omega^*$}
 \end{subfigure}\\ 
 \caption{Stationary solution for $(\sigma,\tau)=(5.0, 0.05)$}
 \label{fig:tau0.05}
\end{figure}

\begin{figure}[H]
 \begin{subfigure}{0.5\columnwidth}
  \centering
  \includegraphics[width=\columnwidth]{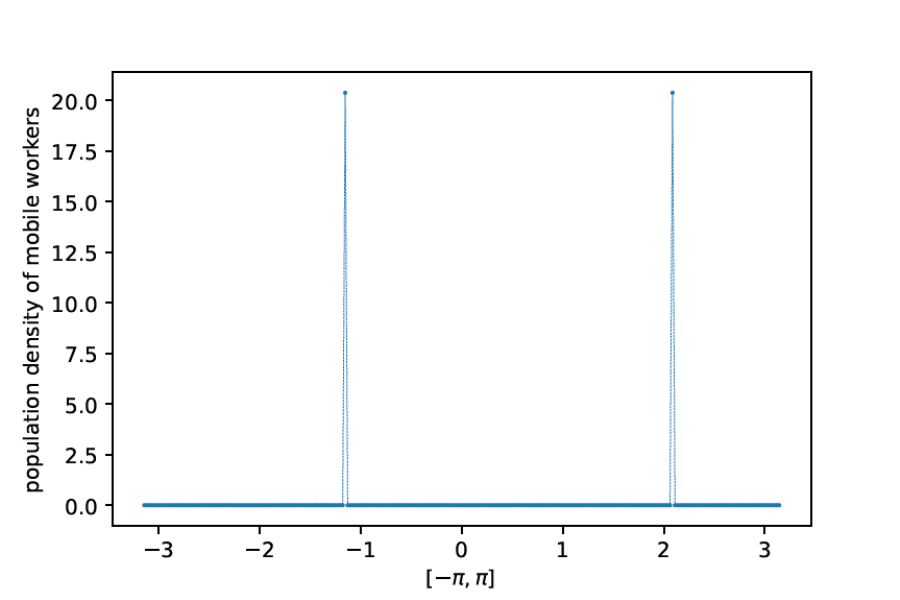}
  \caption{Mobile population density $\lambda^*$}
 \end{subfigure}
 \begin{subfigure}{0.5\columnwidth}
  \centering
  \includegraphics[width=\columnwidth]{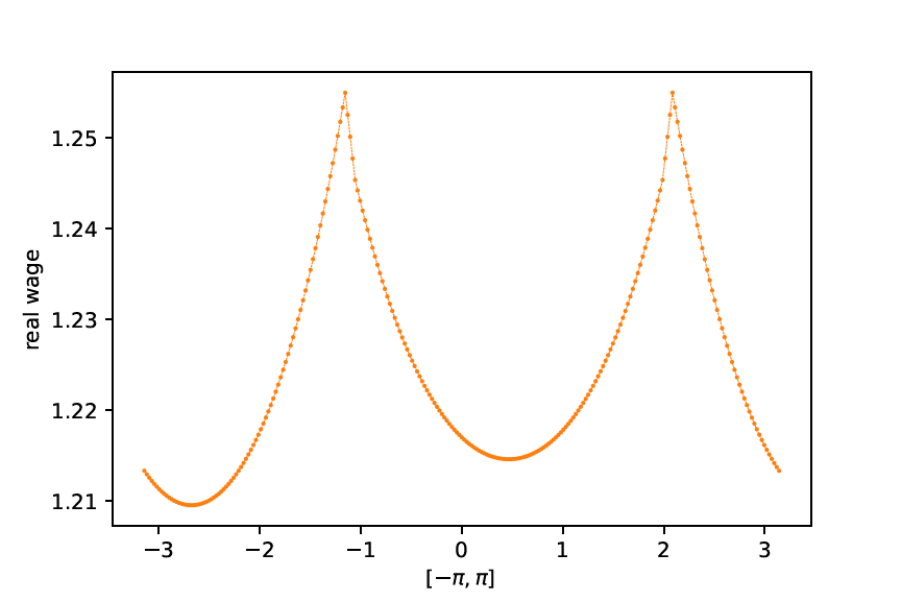}
  \caption{Real wage $\omega^*$}
 \end{subfigure}\\ 
 \caption{Stationary solution for $(\sigma,\tau)=(5.0, 0.1)$}
 \label{fig:tau0.1}
\end{figure}

\begin{figure}[H]
 \begin{subfigure}{0.5\columnwidth}
  \centering
  \includegraphics[width=\columnwidth]{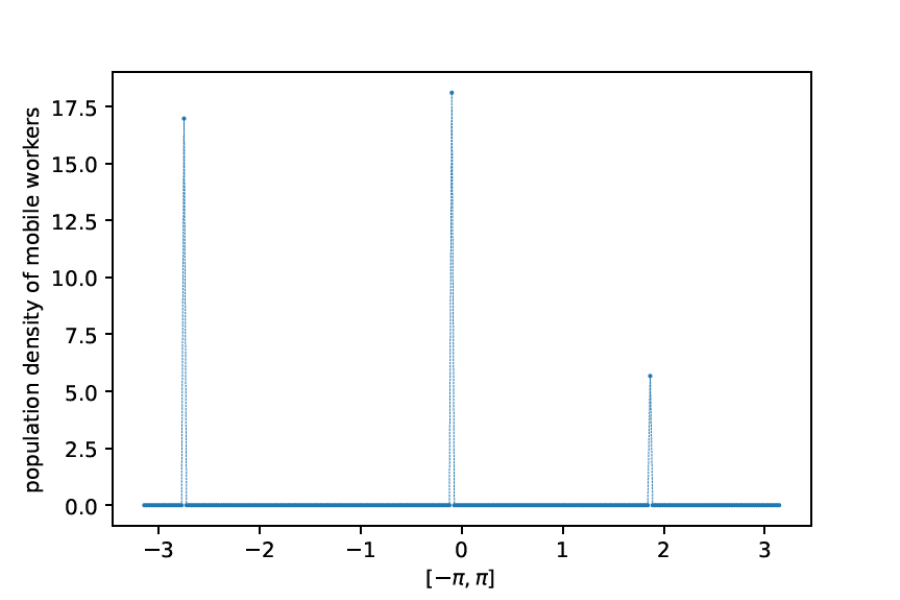}
  \caption{Mobile population density $\lambda^*$}
 \end{subfigure}
 \begin{subfigure}{0.5\columnwidth}
  \centering
  \includegraphics[width=\columnwidth]{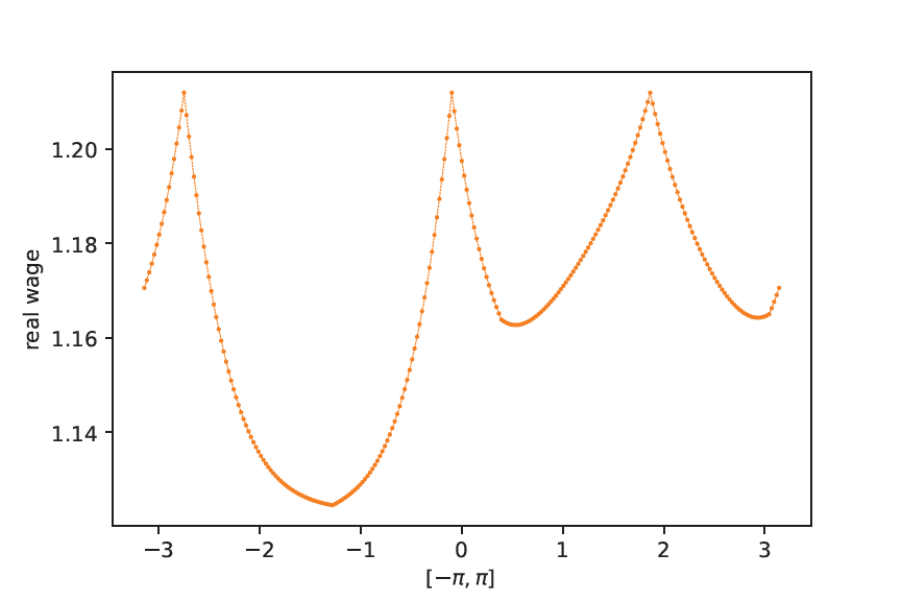}
  \caption{Real wage $\omega^*$}
 \end{subfigure}\\ 
 \caption{Stationary solution for $(\sigma,\tau)=(5.0, 0.2)$}
 \label{fig:tau0.2}
\end{figure}

\begin{figure}[H]
 \begin{subfigure}{0.5\columnwidth}
  \centering
  \includegraphics[width=\columnwidth]{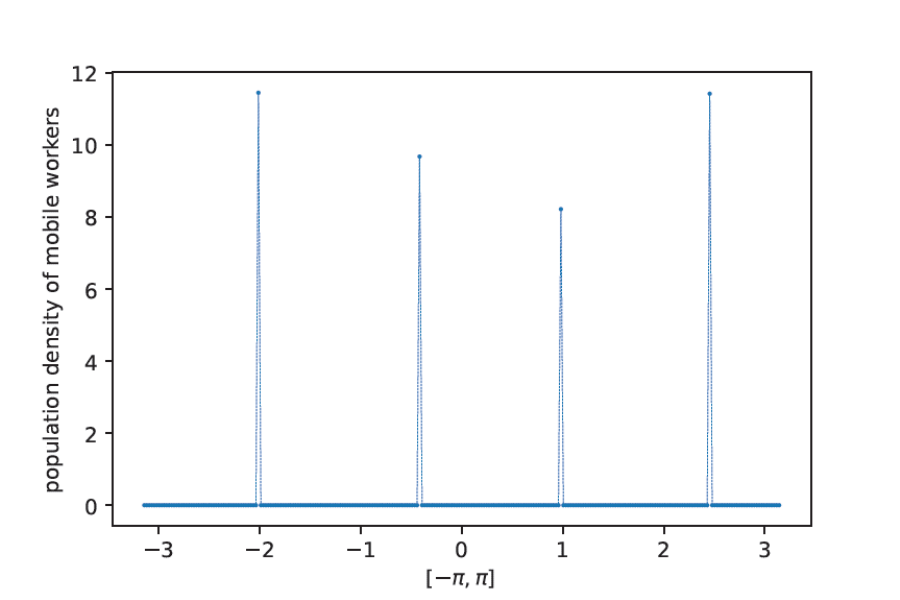}
  \caption{Mobile population density $\lambda^*$}
 \end{subfigure}
 \begin{subfigure}{0.5\columnwidth}
  \centering
  \includegraphics[width=\columnwidth]{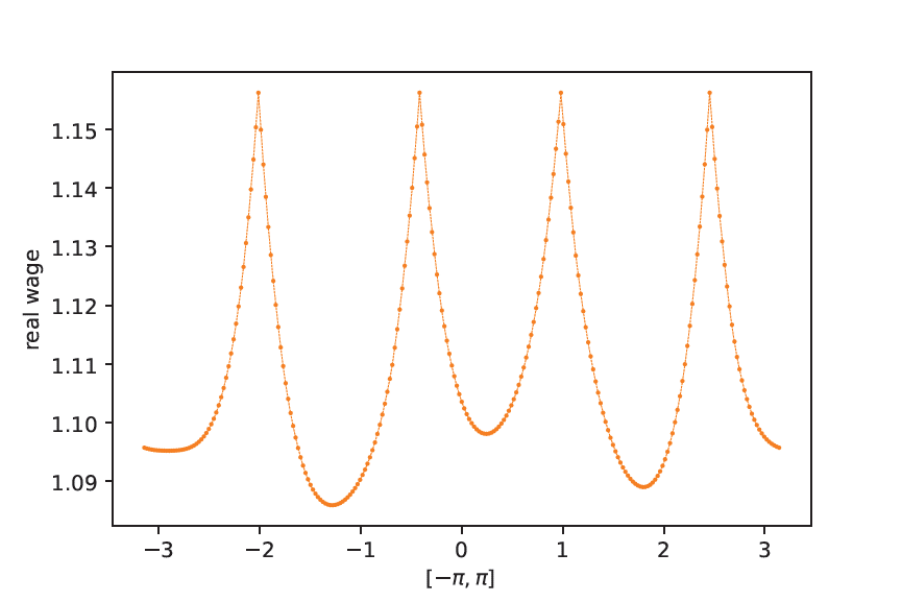}
  \caption{Real wage $\omega^*$}
 \end{subfigure}\\ 
 \caption{Stationary solution for $(\sigma,\tau)=(5.0, 0.3)$}
 \label{fig:tau0.3}
\end{figure}

\begin{figure}[H]
 \begin{subfigure}{0.5\columnwidth}
  \centering
  \includegraphics[width=\columnwidth]{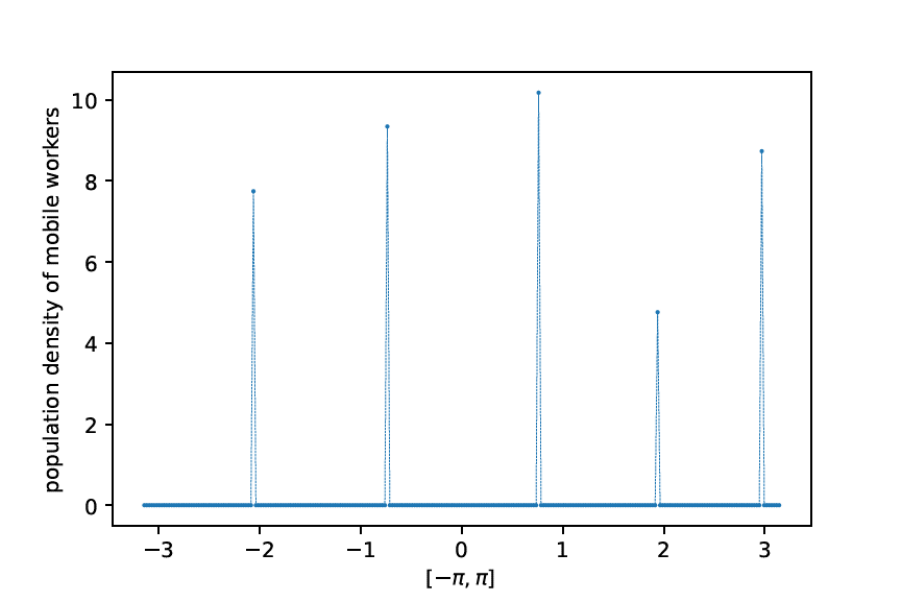}
  \caption{Mobile population density $\lambda^*$}
 \end{subfigure}
 \begin{subfigure}{0.5\columnwidth}
  \centering
  \includegraphics[width=\columnwidth]{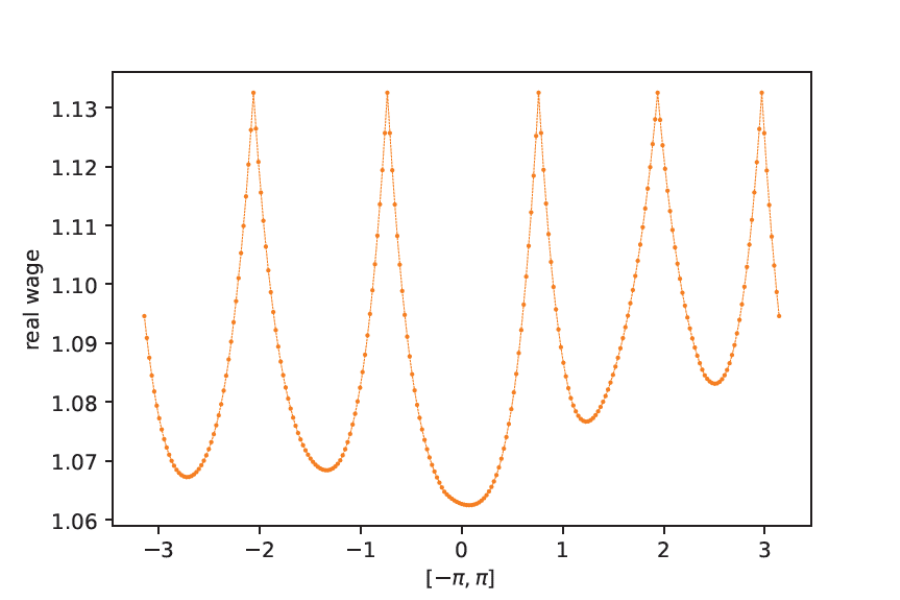}
  \caption{Real wage $\omega^*$}
 \end{subfigure}\\ 
 \caption{Stationary solution for $(\sigma,\tau)=(5.0, 0.35)$}
 \label{fig:tau0.35}
\end{figure}

\begin{figure}[H]
 \begin{subfigure}{0.5\columnwidth}
  \centering
  \includegraphics[width=\columnwidth]{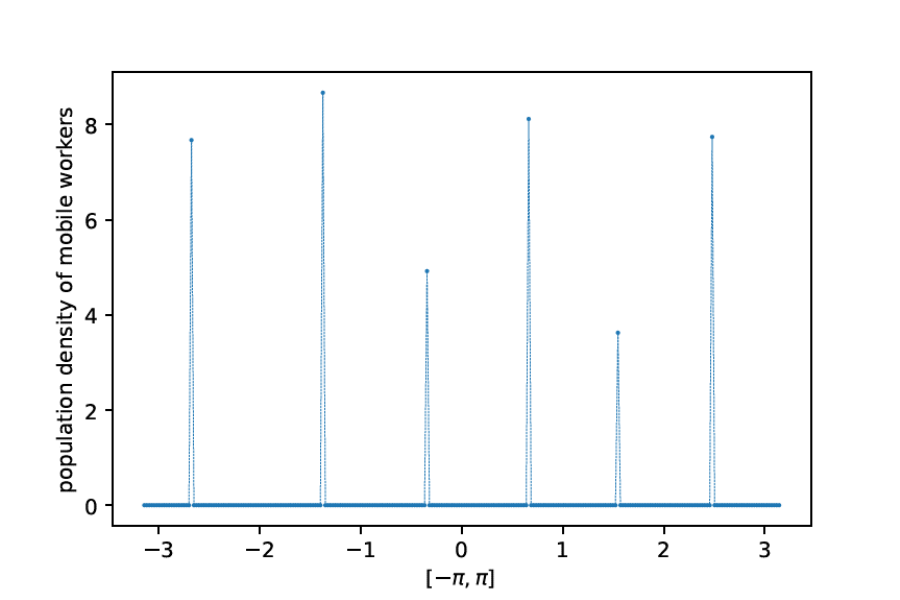}
  \caption{Mobile population density $\lambda^*$}
 \end{subfigure}
 \begin{subfigure}{0.5\columnwidth}
  \centering
  \includegraphics[width=\columnwidth]{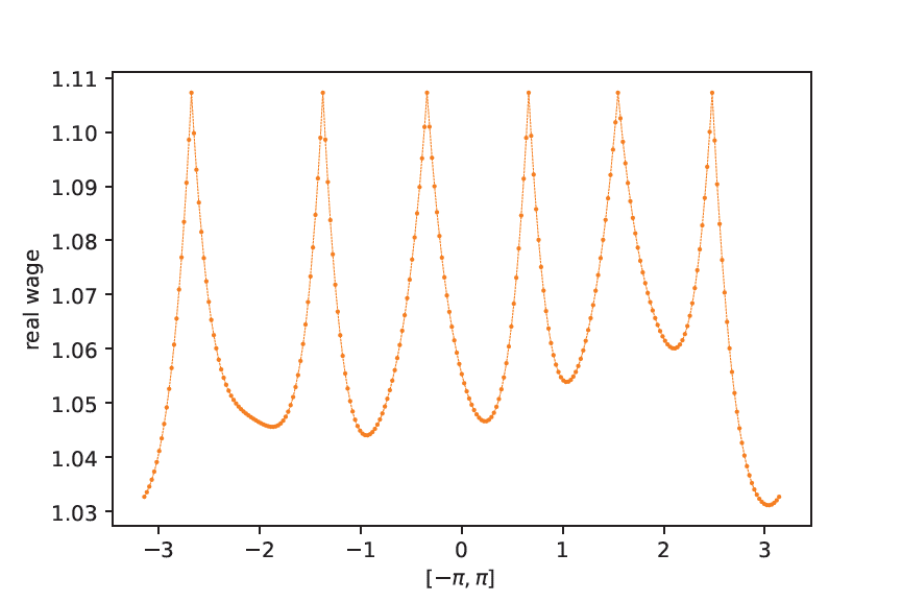}
  \caption{Real wage $\omega^*$}
 \end{subfigure}\\ 
 \caption{Stationary solution for $(\sigma,\tau)=(5.0, 0.42)$}
 \label{fig:tau0.42}
\end{figure}

The stationary states for each value of $\tau$ are shown in Figures \ref{fig:tau0.05}-\ref{fig:tau0.42}. In each figure, the left subfigure shows the approximated population density $\lambda^*(x),~x\in S$, and the right subfigure shows the approximated realwage $\omega^*(x),~x\in S$. Here, we fix $\sigma=5.0$ and vary $\tau$ to $0.05$, $0.1$, $0.2$, $0.3$, $0.35$, and $0.42$. When $\tau$ is sufficiently small, a single spike is formed, and the mobile population density in other regions is almost zero. As the value of $\tau$ increases, the maximum number of the spikes increases, reaching $6$ at $\tau=0.41$. 

\subsection{Change in preference for variety}

Next, we focus on the effects of the preference for variety of consumers on the behavior of the solution of the racetrack model, so we only change $\sigma>1$ here\footnote{We adopt $\tau=0.2$ to have a common setting with the case of varying the value of $\tau$ (Figure \ref{fig:tau0.2}).}.

\begin{figure}[H]
 \begin{subfigure}{0.5\columnwidth}
  \centering
  \includegraphics[width=\columnwidth]{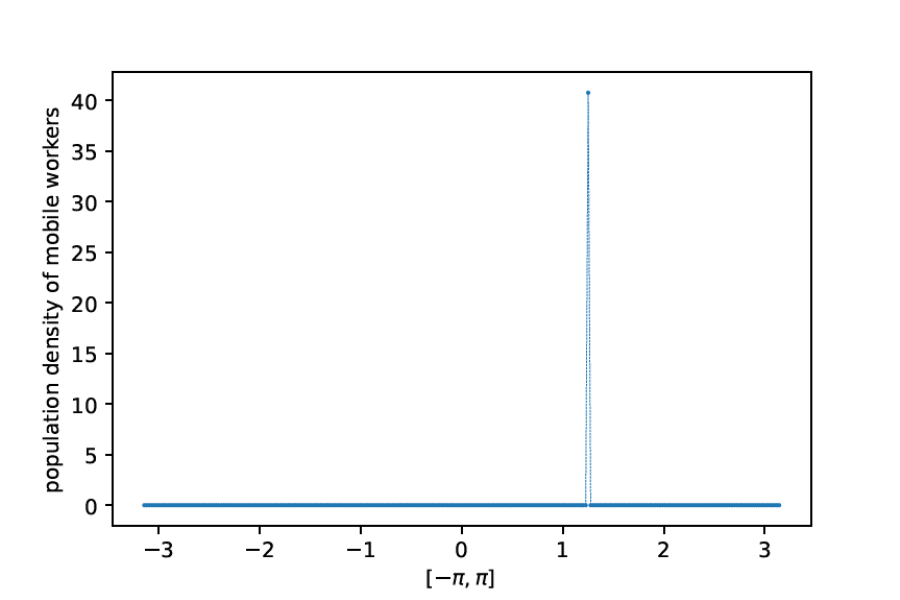}
  \caption{Mobile population density $\lambda^*$}
 \end{subfigure}
 \begin{subfigure}{0.5\columnwidth}
  \centering
  \includegraphics[width=\columnwidth]{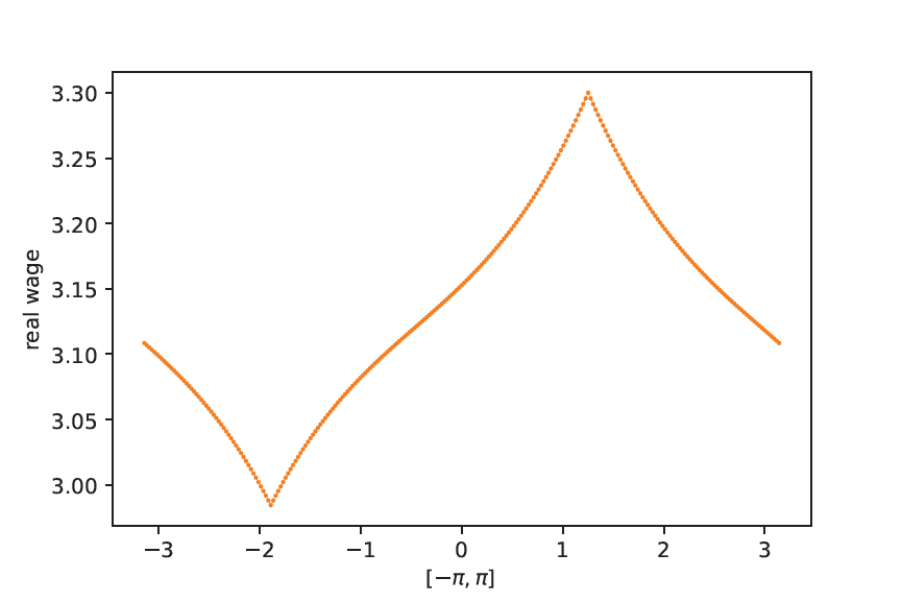}
  \caption{Real wage $\omega^*$}
 \end{subfigure}\\ 
 \caption{Stationary solution for $(\sigma,\tau)=(2.0, 0.2)$}
 \label{fig:sig2}
\end{figure}

\begin{figure}[H]
 \begin{subfigure}{0.5\columnwidth}
  \centering
  \includegraphics[width=\columnwidth]{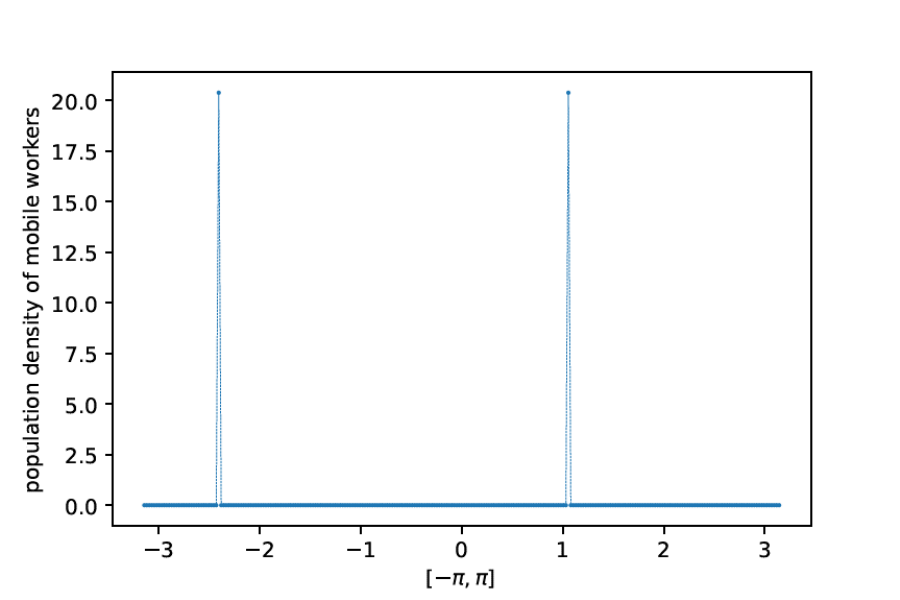}
  \caption{Mobile population density $\lambda^*$}
 \end{subfigure}
 \begin{subfigure}{0.5\columnwidth}
  \centering
  \includegraphics[width=\columnwidth]{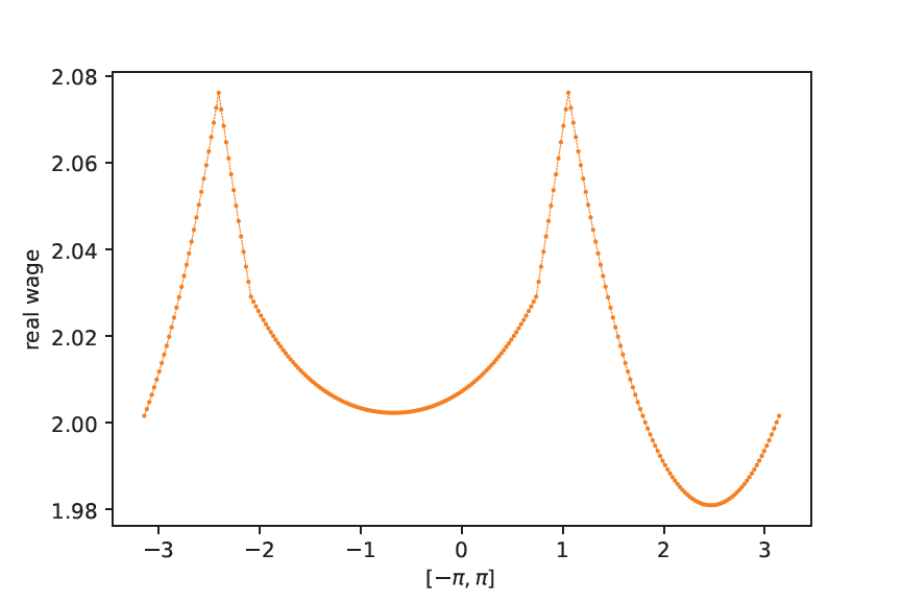}
  \caption{Real wage $\omega^*$}
 \end{subfigure}\\ 
 \caption{Stationary solution for $(\sigma,\tau)=(3.0, 0.2)$}
 \label{fig:sig3}
\end{figure}

\begin{figure}[H]
 \begin{subfigure}{0.5\columnwidth}
  \centering
  \includegraphics[width=\columnwidth]{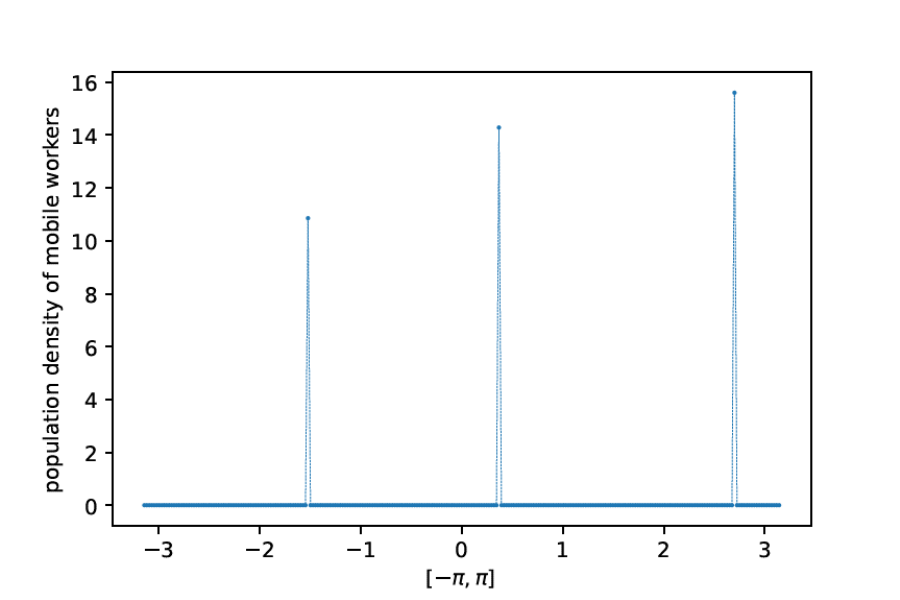}
  \caption{Mobile population density $\lambda^*$}
 \end{subfigure}
 \begin{subfigure}{0.5\columnwidth}
  \centering
  \includegraphics[width=\columnwidth]{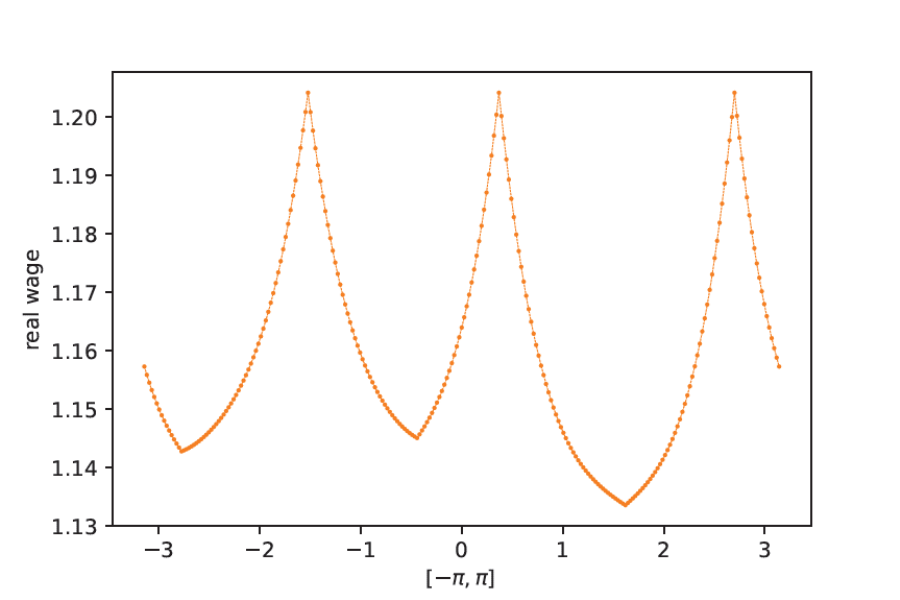}
  \caption{Real wage $\omega^*$}
 \end{subfigure}\\ 
 \caption{Stationary solution for $(\sigma,\tau)=(5.0, 0.2)$}
 \label{fig:sig5}
\end{figure}

\begin{figure}[H]
 \begin{subfigure}{0.5\columnwidth}
  \centering
  \includegraphics[width=\columnwidth]{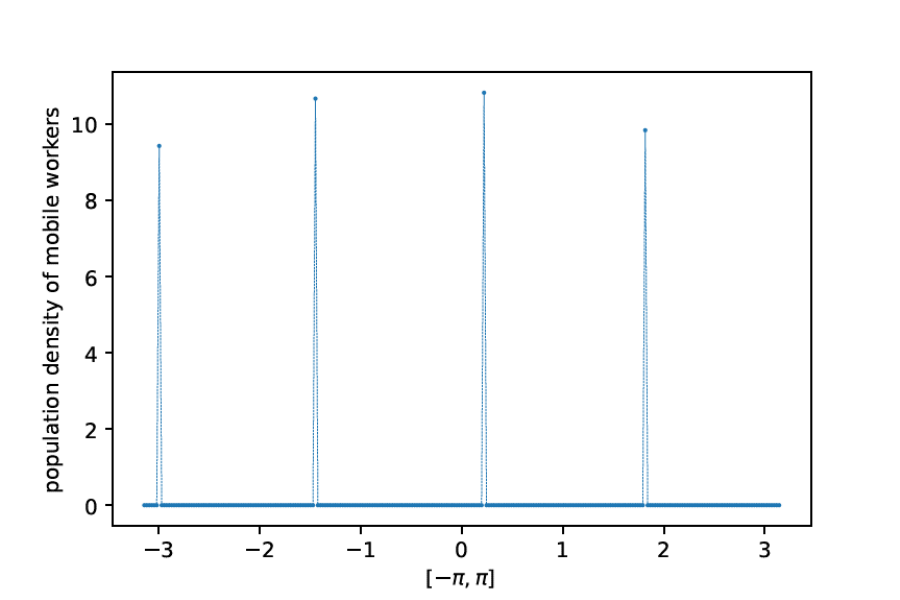}
  \caption{Mobile population density $\lambda^*$}
 \end{subfigure}
 \begin{subfigure}{0.5\columnwidth}
  \centering
  \includegraphics[width=\columnwidth]{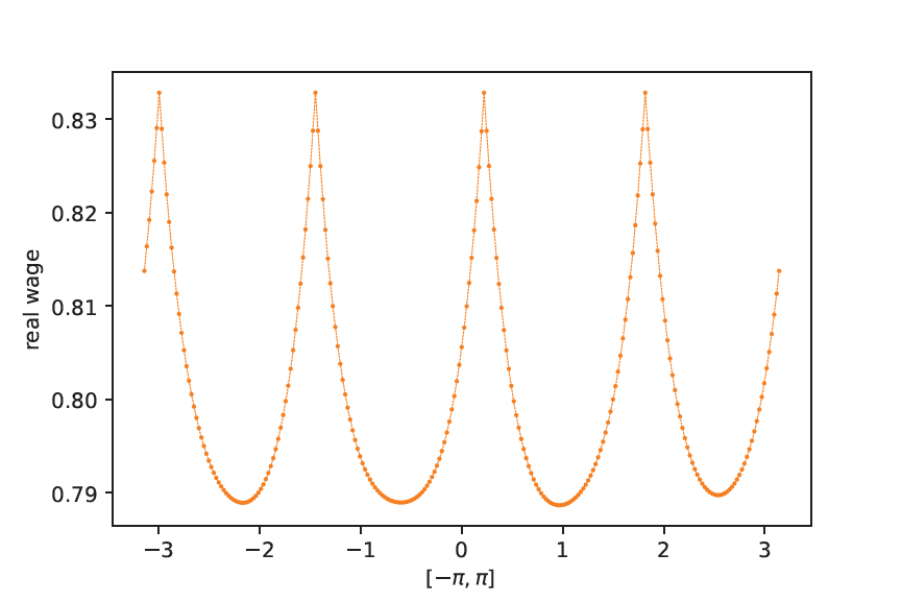}
  \caption{Real wage $\omega^*$}
 \end{subfigure}\\ 
 \caption{Stationary solution for $(\sigma,\tau)=(7.0, 0.2)$}
 \label{fig:sig7}
\end{figure}

\begin{figure}[H]
 \begin{subfigure}{0.5\columnwidth}
  \centering
  \includegraphics[width=\columnwidth]{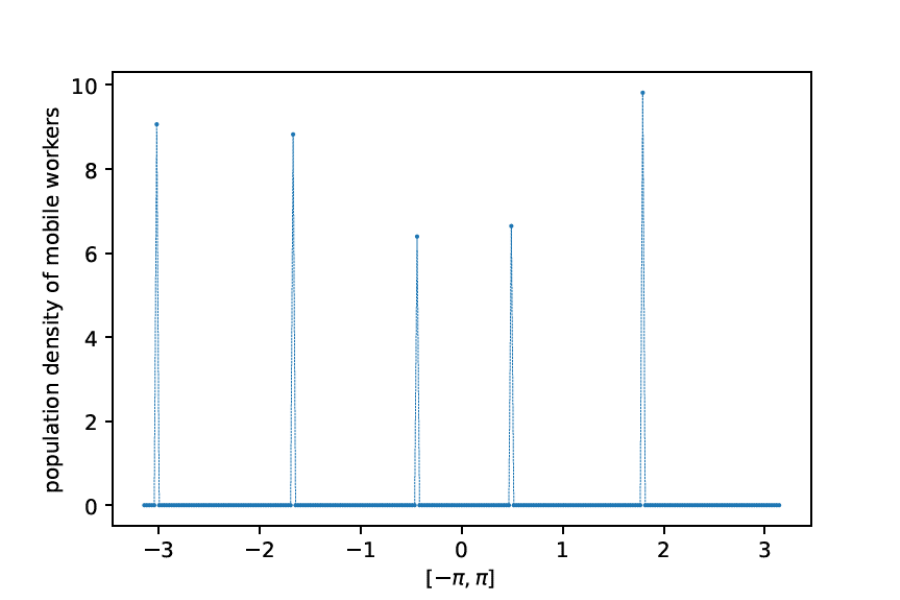}
  \caption{Mobile population density $\lambda^*$}
 \end{subfigure}
 \begin{subfigure}{0.5\columnwidth}
  \centering
  \includegraphics[width=\columnwidth]{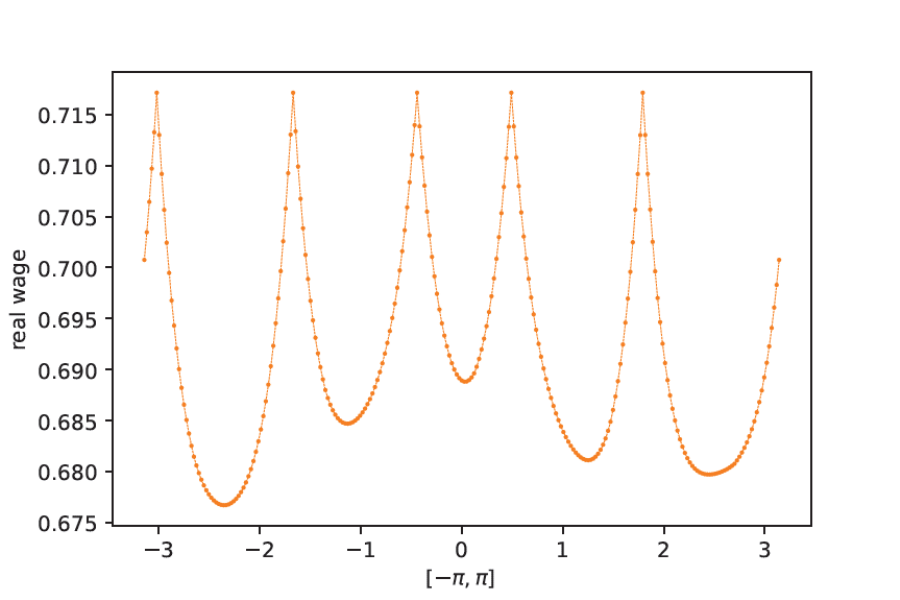}
  \caption{Real wage $\omega^*$}
 \end{subfigure}\\ 
 \caption{Stationary solution for $(\sigma,\tau)=(8.0, 0.2)$}
 \label{fig:sig8}
\end{figure}

\begin{figure}[H]
 \begin{subfigure}{0.5\columnwidth}
  \centering
  \includegraphics[width=\columnwidth]{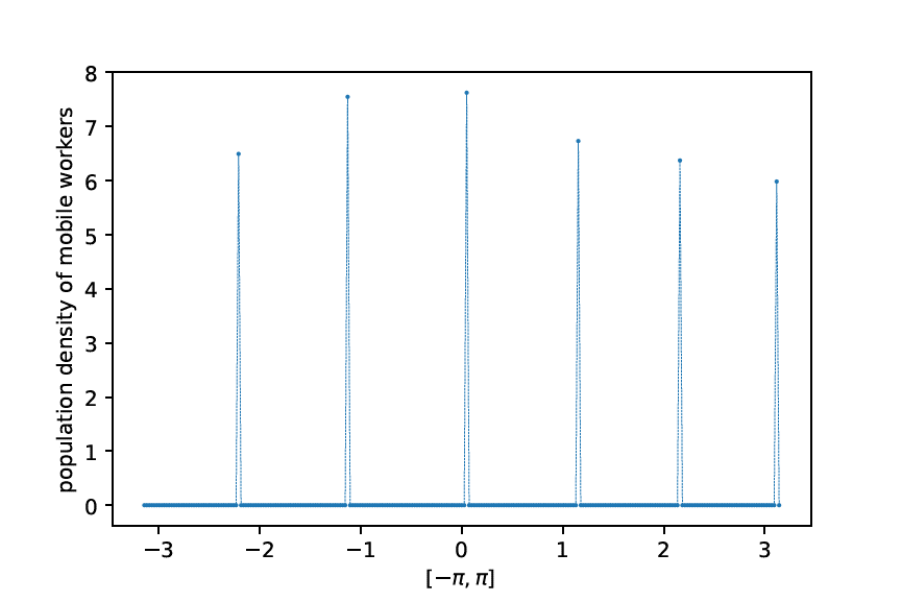}
  \caption{Mobile population density $\lambda^*$}
 \end{subfigure}
 \begin{subfigure}{0.5\columnwidth}
  \centering
  \includegraphics[width=\columnwidth]{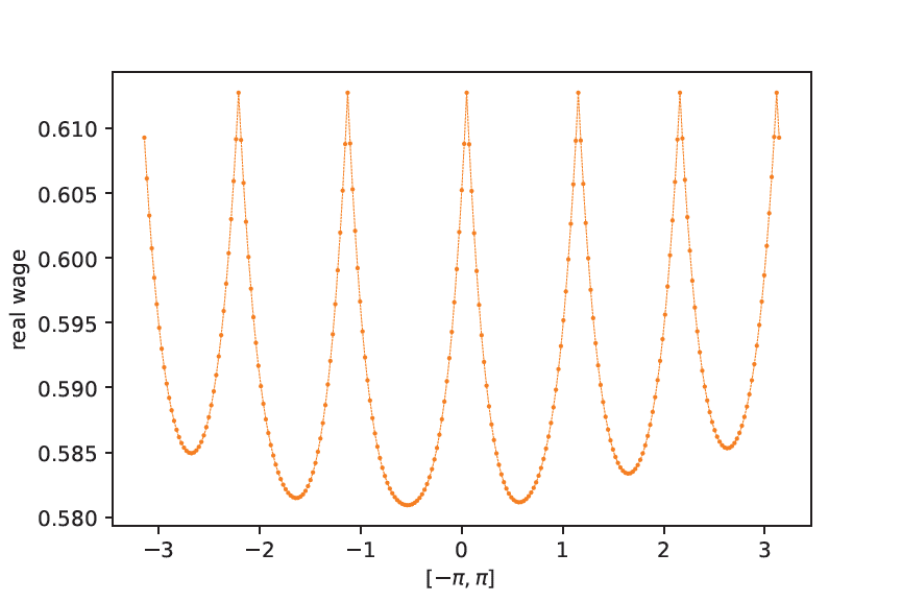}
  \caption{Real wage $\omega^*$}
 \end{subfigure}\\ 
 \caption{Stationary solution for $(\sigma,\tau)=(9.2, 0.2)$}
 \label{fig:sig9.2}
\end{figure}

The stationary states for each $\sigma$ are shown in Figures \ref{fig:sig2}-\ref{fig:sig9.2}. Here, we fix $\tau=0.2$ and vary $\sigma$ to $2.0$, $3.0$, $5.0$, $7.0$, $8.0$, and $9.3$. When the value of $\sigma$ is sufficiently small, a single spike is formed, and the mobile population density in other regions is almost zero. As the value of $\sigma$ increases, the maximum number of the spikes increases, reaching $6$ at $\sigma=9.2$.

\section{Concluding remarks}
In this paper, we consider the extension of the tractable new economic geography models proposed by \citet{Pfl} and \citet{GasCasCorr} to continuous space. We begin with constructing the global solution. Then, we investigate the behavior of the solution of the racetrack model. 

The homogeneous stationary solution is shown to be unstable. Furthermore, it is shown numerically that the destabilized  solution eventually forms peculiar spatial distributions of mobile workers with several spikes that should be called cities. Then, the number of the cities decreases as the preference for variety intensifies or the transport cost decreases. 

The spiky distribution of population means that agglomeration proceeds thoroughly toward points (mathematically, small regions where the measures are zero), which implies that agglomeration is extremely strong in this model. It is not clear whether there is any regularity in the distance between the spikes and the order of the magnitude of them when multiple spikes emerge. From numerical experiments, they seem to depend very sensitively on the initial values.

The model of this paper shares the following assumptions
\begin{itemize}
\item iceberg transport cost of manufactured goods
\item increasing returns and monopolistic competition in manufacturing
\item perfect competition in agriculture
\item exitence of immobile factors
\end{itemize}
with the original racetrack model. The difference is in the assumptions made regarding the utility function and technology of production. The fact that the behavior of the solutions of the model of this paper does not differ qualitatively from that of the original racetrack model  (See \citet{OhtaYagi_point}) suggests that it is the above assumptions that are essentially relevant. This suggests the validity of using the simplified model of this paper in theoretical studies of new economic geography.

\section{Appendix}
This section is devoted to the proofs omitted in the main text. Recall that $T_1$ and $T_2$ are the lower and upper bounds of $T(x,y)$ in \eqref{ulbT}, respectively.

\subsection{Proof  of Theorem \ref{th:locapr}}

\noindent
\begin{lem}\label{lemm:bounds}
The following inequalities
\begin{align}
&\left(\frac{F}{\Lambda+b}\right)^{\frac{1}{\sigma-1}}T_1 \leq \|G\left(\lambda(t)\right)\|_\infty 
\leq \left(\frac{F}{\Lambda-b}\right)^{\frac{1}{\sigma-1}}T_2, \label{bG}\\
&\|w\left(\lambda(t)\right)\|_\infty
\leq \frac{\mu}{\sigma}\left(\frac{T_2}{T_1}\right)^{\sigma-1}\frac{\Phi+\Lambda+b}{\Lambda-b}, \label{bw}\\
&\|\omega\left(\lambda(t)\right)\|_\infty \leq \frac{\mu}{\sigma}\left(\frac{T_2}{T_1}\right)^{\sigma-1}\frac{\Phi+\Lambda+b}{\Lambda-b}\nonumber\\
&\hspace{14mm} + \mu \max\left\{\left|\ln \left[\left(\frac{F}{\Lambda-b}\right)^{\frac{1}{\sigma-1}}T_1\right]\right|, \left|\ln\left[\left(\frac{F}{\Lambda-b}\right)^{\frac{1}{\sigma-1}}T_2\right]\right|\right\} \label{bomega}
\end{align}
hold for the operators \eqref{opG}, \eqref{opw}, and \eqref{opom}.
\end{lem}

\vspace{3mm}
\noindent
(Proof of Lemma \ref{lemm:bounds})\\

Let us first observe that any $\lambda$ such that $\left\|\lambda(t)-\lambda_0\right\|_{L^1}\leq b$ satisfies 
\begin{equation}\label{lamestim}
0<\Lambda-b\leq \left\|\lambda(t)\right\|_{L^1}\leq \Lambda+b.
\end{equation}
Then, from \eqref{opG} and \eqref{lamestim}, it follows that
\begin{equation*}
\begin{aligned}
F^{\frac{1}{\sigma-1}}T_1\left(\Lambda+b\right)^{\frac{1}{1-\sigma}}\leq \left|G(\lambda(t))(x)\right|\leq F^{\frac{1}{\sigma-1}}T_2\left(\Lambda-b\right)^{\frac{1}{1-\sigma}},
\end{aligned}
\end{equation*}
which immediately yield \eqref{bG}. It follows from \eqref{bG}, \eqref{opw}, and  \eqref{lamestim} that
\[
\begin{aligned}
\left|w(\lambda(t))(x)\right|
&\leq \frac{\mu}{\sigma F}T_1^{1-\sigma}\left[\left(\frac{F}{\Lambda-b}\right)^{\frac{1}{\sigma-1}}T_2\right]^{\sigma-1}\left(\Phi+\Lambda+b\right),
\end{aligned}
\]
which immediately yields \eqref{bw}. It follows from \eqref{bG} and \eqref{bw} that
\[
\begin{aligned}
\left|\omega(x)\right| &\leq
\left|w(x)\right| + \mu\left|\ln G(x)\right|,\\
&\leq 
\frac{\mu}{\sigma}\left(\frac{T_2}{T_1}\right)^{\sigma-1}\frac{\Phi+\Lambda+b}{\Lambda-b}\\
&\hspace{5mm}+ \mu \max\left\{\left|\ln \left[\left(\frac{F}{\Lambda-b}\right)^{\frac{1}{\sigma-1}}T_1\right]\right|, \left|\ln\left[\left(\frac{F}{\Lambda-b}\right)^{\frac{1}{\sigma-1}}T_2\right]\right|\right\},
\end{aligned}
\]
which immediately gives \eqref{bomega}.

\begin{flushright}
(End of Proof of Lemma \ref{lemm:bounds})
\end{flushright}

By \eqref{opPsi} and \eqref{lamestim}, we have that
\[
\begin{aligned}
\left\|\Psi\left(\lambda(t)\right)\right\|_{L^1}
&= \gamma \int_{\cl M}\left|\left[\omega(\lambda(t))(x)-\frac{1}{\Lambda}\int_{\cl M}\omega(\lambda(t))(y)\lambda(y) dy\right]\lambda(x)\right|dx\\
&\leq \gamma \left\|\omega(\lambda(t))\right\|_\infty \left(\Lambda+b\right)\left(1+\frac{\Lambda+b}{\Lambda}\right).
\end{aligned}
\]
Together with \eqref{bomega}, this completes the proof.  \qed

\subsection{Proof of Theorem \ref{th:lip}}

\noindent

Firstly, we show that $G$ is Lipschitz continuous. For $\lambda_1,\lambda_2\in Q$, let us define
\[
g_i(x) := \frac{1}{F}\int_{\cl M}\left|\lambda_i(y)\right|T(x,y)^{1-\sigma}dy,~i=1,2
\]
so that $G\left(\lambda_i\right)(x) = g_i(x)^{\frac{1}{1-\sigma}}$. Then, based on the mean-value theorem in a Banach space, we obtain
\begin{equation}\label{G1G2g1g2}
\left\|G\left(\lambda_1\right)-G\left(\lambda_2\right)\right\|_\infty \leq \cl{C}\left\|g_1-g_2\right\|_\infty,
\end{equation}
where $\cl{C}=\frac{F^{\frac{\sigma}{1-\sigma}}T_2^{\sigma}(\Lambda-b)^{\frac{\sigma}{1-\sigma}}}{\sigma-1}$. Then, we see that
\begin{align}
\left\|g_1-g_2\right\|_\infty
&=\max_{x\in \cl{M}}\left|\frac{1}{F}\int_{\cl M}\left(|\lambda_1(y)|-|\lambda_2(y)|\right)T(x,y)^{1-\sigma}dy\right| \nonumber\\
&\leq \frac{T_1^{1-\sigma}}{F}\int_{\cl M}\left|\left|\lambda_1(y)\right|-\left|\lambda_2(y)\right|\right|dy\nonumber \\
&\leq \frac{T_1^{1-\sigma}}{F}\int_{\cl M}\left|\lambda_1(y)-\lambda_2(y)\right|dy
= \frac{T_1^{1-\sigma}}{F} \left\|\lambda_1-\lambda_2\right\|_{L^1}. \label{g1g2}
\end{align}
From \eqref{G1G2g1g2} and \eqref{g1g2}, we have
\begin{equation}\label{LG}
\left\|G\left(\lambda_1\right)-G\left(\lambda_2\right)\right\|_\infty \leq 
\cl{L}_G \left\|\lambda_1-\lambda_2\right\|_\infty,
\end{equation}
where $\cl{L}_G>0$ is a constant.

Secondly, we show that $w$ is Lipschitz continuous. 
\begin{align}
\|w(\lambda_1)-w(\lambda_2)\|_\infty &= \max_{x\in \cl{M}}\left|w(\lambda_1)(x)-w(\lambda_2)(x)\right| \nonumber\\
&= \max_{x\in \cl{M}}\left|\frac{\mu}{\sigma F}\int_{\cl M}\left(\phi(y)+\left|\lambda_1(y)\right|\right)G(\lambda_1)(y)^{\sigma-1}T(x,y)^{1-\sigma}dy\right. \nonumber\\
&\hspace{5mm}\left.-\frac{\mu}{\sigma F}\int_{\cl M}\left(\phi(y)+\left|\lambda_2(y)\right|\right)G(\lambda_2)(y)^{\sigma-1}T(x,y)^{1-\sigma}dy\right| \nonumber\\
&\leq \frac{\mu}{\sigma F}T_1^{1-\sigma}\left(\Phi+\left\|\lambda_1\right\|_{L^1}\right)\left\|G(\lambda_1)^{\sigma-1}-G(\lambda_2)^{\sigma-1}\right\|_\infty \nonumber\\
&\hspace{20mm} + \frac{\mu}{\sigma F}T_1^{1-\sigma}\left\|G^{\sigma-1}\right\|_\infty\left\|\lambda_1-\lambda_2\right\|_{L^1} \nonumber \\
&\leq \frac{\mu}{\sigma F}T_1^{1-\sigma}\left(\Phi+\Lambda+b\right)\left\|G(\lambda_1)^{\sigma-1}-G(\lambda_2)^{\sigma-1}\right\|_\infty \nonumber\\
&\hspace{20mm} + \frac{\mu}{\sigma}\left(\frac{T_2}{T_1}\right)^{\sigma-1}\frac{1}{\Lambda-b}\left\|\lambda_1-\lambda_2\right\|_{L^1} \label{w1w2l1l2}
\end{align}
Here, \eqref{bG} and \eqref{lamestim} are used in the last deformation. Then, by the mean-value theorem and \eqref{bG}, we obtain
\begin{equation}\label{G1G2sigma1G1G2}
\left\|G(\lambda_1)^{\sigma-1}-G(\lambda_2)^{\sigma-1}\right\|_\infty
\leq \cl{C}\left\|G(\lambda_1)-G(\lambda_2)\right\|_\infty,
\end{equation}
where
\[
\cl{C} = \left\{
\begin{aligned}
&(\sigma-1)\left(\frac{F}{\Lambda-b}\right)^{\frac{\sigma-2}{\sigma-1}}T_2^{\sigma-2},~~~{\rm if}~\sigma-2\geq 0\\
&(\sigma-1)\left(\frac{F}{\Lambda-b}\right)^{\frac{\sigma-2}{\sigma-1}}T_1^{\sigma-2},~~~{\rm if}~\sigma-2 <0.
\end{aligned}
\right.
\]
From \eqref{w1w2l1l2} and \eqref{G1G2sigma1G1G2}, we obtain
\begin{equation}\label{Lw}
\left\|w(\lambda_1)-w(\lambda_2)\right\|_\infty \leq \cl{L}_w \left\|\lambda_1-\lambda_2\right\|_\infty,
\end{equation}
where $\cl{L}_w>0$ is a constant.

Thirdly, we show that $\omega$ is Lipschitz continuous. By the mean-value theorem and \eqref{bG}, we obtain
\begin{equation}\label{lnG1lnG2mv}
\left\|\ln G(\lambda_1)-\ln G(\lambda_2)\right\|_\infty \leq 
\left(\frac{\Lambda-b}{F}\right)^{\frac{1}{\sigma-1}}T_1\left\|G(\lambda_1) - G(\lambda_2)\right\|_\infty
\end{equation}
It follows from \eqref{Lw} and \eqref{lnG1lnG2mv} that
\begin{align}
\left\|\omega(\lambda_1)-\omega(\lambda_2)\right\|_\infty
&\leq \left\|w(\lambda_1)-w(\lambda_2)\right\|_\infty + \mu\left\|\ln G(\lambda_1)- \ln G(\lambda_2)\right\|_\infty \nonumber\\
&\leq 
\cl{L}_{\omega} \|\lambda_1-\lambda_2\|_{L^1} \label{Lomega}
\end{align}
where $\cl{L}_\omega>0$ is a constant.

We are now able to show the Lipschitz continuity of $\Psi(\lambda)$. By \eqref{bomega} and \eqref{lamestim}, we see that
\begin{align}
\left\|\Psi(\lambda_1)-\Psi(\lambda_2)\right\|_{L^1}
&= \gamma\int_{\cl M}\left|\omega(\lambda_1)(x)\lambda_1(x) - \frac{1}{\Lambda}\int_\cl{M}\omega(\lambda_1)(y)\lambda_1(y)dy\cdot\lambda_1(x)\right. \nonumber\\
&\hspace{10mm}\left.-\omega(\lambda_2)(x)\lambda_2(x) + \frac{1}{\Lambda}\int_\cl{M}\omega(\lambda_2)(y)\lambda_2(y)dy\cdot\lambda_2(x)\right|dx \nonumber\\
&\leq 2\gamma\left\|\omega\right\|_\infty\left\|\lambda_1-\lambda_2\right\|_{L^1} \nonumber\\
&\hspace{15mm}+ \gamma(\Lambda+b)^2\left\|\omega(\lambda_1)-\omega(\lambda_2)\right\|_\infty \label{Lpsiom12}
\end{align}
Thus, \eqref{Lomega} and \eqref{Lpsiom12} complete the proof.

\qed

\subsection{Proof of Theorem \ref{th:apr}}
For any $\lambda\in L^1_\Lambda$, discussion similar to that in the proof of Theorem \ref{th:locapr} but now with $\left\|\lambda\right\|_{L^1}=\Lambda$ instead of \eqref{lamestim} completes the proof. 

\qed

\bibliographystyle{aer}
\ifx\undefined\bysame
\newcommand{\bysame}{\leavevmode\hbox to\leftmargin{\hrulefill\,\,}}
\fi

\end{document}